\documentclass[useAMS, usenatbib]{mn2e}

\pdfoutput=1 

\newcommand{\aap}{A\&A}
\newcommand{\apj}{ApJ}
\newcommand{\mnras}{MNRAS}
\newcommand{\prd}{Phys. Rev. D}
\newcommand{\sovast}{Soviet Astronomy}
\newcommand{\nat}{Nature}
\newcommand{\jcap}{Journal of Cosmology as Astroparticle Physics}
\newcommand{\araa}{Annual Review of Astronomy as Astrophysics}

\input{epsf}

\usepackage{amssymb, amstext, amsmath, epsf, graphicx, graphics, epsfig, wrapfig, listings, float, esint}
\usepackage{hyperref, natbib, ifthen}

\usepackage{xcolor}
\usepackage[normalem]{ulem} 
\newcommand\hl{\bgroup\markoverwith
  {\textcolor{yellow}{\rule[-.5ex]{2pt}{2.5ex}}}\ULon}

\begin{document}

\title[Paper 1]{Multi-stream portrait of the Cosmic web}

\author[Ramachandra \& Shandarin]
	{Nesar S. Ramachandra, \thanks{E-mail: nesar@ku.edu} 
	Sergei F. Shandarin, \\
	Department of Physics and Astronomy, University of Kansas, Lawrence, KS 66045}

 \maketitle
\begin{abstract}
We report the results of the first study of the multi-stream environment of dark matter haloes in cosmological N-body simulations
in the $\Lambda$CDM cosmology. The full dynamical state of  dark matter can be described as a three-dimensional sub-manifold
in six-dimensional phase space - the dark matter sheet. In our study we use a Lagrangian sub-manifold ${\bf x} = {\bf x}({\bf q},t)$ 
(where {\bf x} and {\bf q} are co-moving Eulerian and Lagrangian coordinates respectively), which 
is dynamically  equivalent to the dark matter sheet but is more convenient for numerical analysis.
Our major results can be summarized as follows. At the resolution of the simulation i.e. without additional
smoothing, the cosmic web represents a hierarchical structure: each halo is embedded in the filamentary framework of the web predominantly 
at the filament crossings, 
and each filament is embedded in the wall like fabric of the web at the wall crossings. 
Locally, each halo or sub-halo is a peak in the number of streams field. The number of streams in the neighbouring  
filaments is higher than in the neighbouring walls. The walls are regions where number of streams
 is equal to three or a few. Voids are uniquely defined by the local condition requiring to be a single-stream flow region.
  The shells of streams 
 around haloes are  quite thin and the closest void region is typically within one and a half FOF radius 
 from the center of the halo.
\end{abstract}


\section{Introduction} \label{sec:intro}

The problem of objective identification and quantitative characterization of anisotropic structures 
in the distribution of galaxies in space emerged after the first evidences of their existence
(see the review by \citealt{Oort:83} and the references therein). The first theoretical model predicting highly anisotropic concentrations in the mass distribution coming into existence at the non-linear stage of gravitational instability is known as the Zeldovich Approximation (the ZA) (\citealt{Zeldovich_aap:70}, for further developments see also \citealt{Shandarin_Zeldovich:89} and the references therein). The ZA predicted the formation of  `pancakes'  also known as the walls in the currently popular jargon. The later development
of the model by \citet{Arnold_etal:82} predicted the formation of filaments along with the pancakes. \citet{Klypin_Shandarin:83} and \citet{Shandarin_Klypin:84} demonstrated that the filaments emerge in the cosmological N-body simulation in three-dimensional space.
However, they failed to identify the pancakes at $z=0$. Both the existence of filaments connecting compact clumps of matter and absence of pancakes were confirmed by \citet{Frenk_etal:83}. Puzzled by the absence of the pancakes \citet{Klypin_Shandarin:83}
speculated that insufficient mass resolution of the simulation was the cause  of the negative outcome. 
This has been unambiguously confirmed by recent simulations using a better numerical technique of computing a density field from the particle
coordinates in cosmological N-body simulations (~\citealt{Shandarin_etal:12} and ~\citealt{Abel_etal:12}). \citealt{Klypin_Shandarin:83} 
also stressed that the most of filaments are  incorporated in 
`a single three-dimensional web structure'\footnote{The term `cosmic web' was coined by  \citet{Bond_etal:96}.}. 
They admitted that
their simulation did not allow them to confirm the existence of pancakes between the filaments predicted by the ZA \citep{Arnold_etal:82}.
  
Although the four archetypical elements of the cosmic web:  voids, walls/pancakes, filaments and haloes 
were predicted by ZA and confirmed in cosmological N-body simulation their identification and quantitative characterization
remains under vigorous debate
(see e.g. \citealt{Colberg_etal:08}, \citealt{Elahi_etal:13}, \citealt{Knebe_etal:13}, \citealt{Hoffmann_etal:14}).
The dark matter haloes are arguably the easiest objects to identify in N-body simulations. They can also be reliably associated with
observed objects like galaxies and clusters of galaxies.  But even in this case \citet{Knebe_etal:13} refer to almost thirty different
halo finders suggested after 2000. Identifying filaments and pancakes/walls is far more controversial in both N-body simulations 
and galaxy catalogues. For instance, even estimating the global parameters of the  web  in N-body simulation such as the fractions of volume and mass
in voids, walls/pancakes, filaments and haloes produced quite different results. The estimates of volume 
fractions of voids range from 13  to 86\%  (\citealt{Cautun_etal:14}, \citealt{Flack_Neyrinck:14}, \citealt{Forero_etal:09} , \citealt{Hahn_etal:07} , \citealt{Aragon_etal:10}). Similar estimates for walls/pancakes, filaments 
and haloes  are respectively 5-56\% 2-26\% and 0.1-1\%.  Estimates of the mass content vary in large ranges as well.

Large differences in the estimates of volume and mass fractions made by different groups are not surprising if we recognize
considerable differences in the definitions of the components of the cosmic web and numerical methods used in the estimates. Without trying to provide an exhaustive review of all definitions and techniques used for the quantitative morphological 
analysis of the web  we just briefly describe a few approaches in order to illustrate how different they could be. 
Some groups study the web morphology using only coordinates of simulation particles, while others use the particle velocities too. 
Transforming data from  point sets to the density and other fields on a grid is also often used
because fields allow to use a variety of mathematical techniques not available for the particle sets.
However this step can be done by a variety of methods some as simple as cloud-in-cell (CIC), or more complicated as smoothed-particle hydrodynamics (SPH), 
or using Voronoi and Delaunay tessellations
as in the Delaunay Tessellation Field Estimator (DTFE) method (~\citealt{Weygaert_Schaap:09}, \citealt{Cautun_etal:14}).  
Recently a new method called a discrete persistent structure extractor (DisPerSE \citealt{Sousbie:11},  \citealt{Sousbie_etal:11} allowing to identify haloes and other components of the web directly from the particles 
has been designed.  This method can be applied to the galaxy catalogues, for instance \citet{Sousbie_etal:11} applied it to SDSS catalogue and extracted the filaments (which are made available online).

An obvious advantage of methods based on particle coordinates, both based on the density field and directly on particle coordinates, is their applicability to redshift catalogues. 
The redshift catalogues like SDSS and 2dF provide only two angular coordinates and distances in redshift space. 
But cosmological N-body dark matter simulations provide the full dynamical information in six-dimensional
phase space. This additional information is very valuable providing a greater opportunity for understanding the 
physics of the web and developing a better theory of the web.

Dark matter distribution in phase space is highly degenerate because it is cold. 
Practically, it occupies a three-dimensional sub-manifold in six-dimensional phase space.
 In the linear regime, the dark matter sub-manifold is a single-valued function of Eulerian coordinates 
 which means that at each point the dark matter is represented by a single stream flow. As the density perturbations in dark matter grow with time the number of streams jumps to three at the regions of
shell crossing. Then five stream regions emerge inside of the three stream regions and so on.
Number of streams remains an odd integer in generic points.   The corresponding parts of the three-dimensional dark matter sub-manifold 
form complicated folds in six-dimensional  phase space.   

The regions with multi-stream flow constitute the web while the regions with only one stream form
voids (\citealt{Shandarin:11}, ~\citealt{Shandarin_etal:12}, ~\citealt{Abel_etal:12},~\citealt{Flack_etal:12}).
This definition of voids states that in a given N-body simulation, no haloes can be formed before the first
shell crossings have occurred and the smallest haloes cannot be smaller than the mass corresponding 
to the small scale cut-off in the initial power spectrum regardless of the cause of the cut-off:  physical or due to numerical
limitations (see e.g. \citealt{Angulo_etal:13}).  This definition of voids is physical by nature and thus has no  free parameters. In addition, it does not speculate on the sub-grid physical processes.
The first three-stream flow regions are similar to the pancakes in the ZA.
They quickly grow and merge into a complicated three-dimensional structure; filaments making the framework of the web manifest 
themselves at the pancake crossings,  and haloes emerge at the filament crossings. At later times  different parts of the web participating
in the large-scale motion overlap  which increases the web complexity further.

Using the full six-dimensional information allows one to generate new fields which  provide 
additional useful information about the evolution and morphology of the  web. 
One of them is a multi-stream field in Eulerian space, which will be the focus of this paper. 
Another example is the flip-flop field in Lagrangian space. In cosmological context it was first used in the ZA.
\citet{Vogelsberger_etal:11} used it in a study of multi-stream structure of galaxy size haloes.
\citet{Shandarin_Medvedev:14} applied it for identifying subhaloes  in dark matter haloes.
A similar although somewhat simplistic realization of this idea has been revealed in the ORIGAMI method
used for the analysis of the web ( \citealt{Flack_etal:12},  \citealt{Flack_Neyrinck:14} ).
 Although these fields cannot be used directly on observational data because the full phase-space information
 is not available, they provide much deeper insight into non-linear clustering of collision-less dark matter
and reveal new features of the web.


In order to compute the multi-stream field we will use the tessellation scheme described in
 \citet{Shandarin_etal:12} which is  also briefly discussed in Section \ref{subsec:method}.
  Using this methodology on the entire simulation box, we discuss the global behaviour of the number of streams in the cosmic web in Section \ref{sec:global}. 
  The tessellation technique we have utilized can be used to find multi-stream fields in smaller Eulerian boxes with very high resolution too. In Section \ref{sec:local} we study the local behaviour of multi-streams flows in regions around haloes the detected using friends-of-friends (FOF) technique.

\section{The simulation}
\label{sec:simulation}

We have utilized the data from cosmological N-body simulations by Gadget-2 \citep{Springel:05} for 100 $h^{-1}$ Mpc  and 200 $h^{-1}$ Mpc box 
sizes with $128^3$, $256^3$ and $512^3$ grids. Each particle is between  $10^9 - 10^{12} M_{\odot}$. 
The initial conditions and cosmological parameters are consistent with the Planck cosmology. We utilize the initial Lagrangian box and do a three-dimensional mapping onto corresponding evolved simulations. In addition, for local multi-stream analyses around haloes, we have utilized halo catalogues for each of these simulation boxes. These haloes are detected using FOF method considering objects with more than 20 particles found at linking length, $ b= 0.2$.  
\section{Multi-stream field calculation}
\label{subsec:method}

Phase space tessellation considers the dynamics of the particles similar to that of a standard N-body code. 
However the particles are nodes of the tessellation, and are just massless tracers of the flow.
Assuming that the uniform state is modelled by a simple rectangular grid, the particles are the nodes of the grid.
Each elementary cube of the grid is tessellated by five tetrahedra (\citealt{Shandarin_etal:12}\footnote{For the description of an alternative type of the tessellation see \citet{Abel_etal:12}.}) of which the vertices are the
vertices of the cube. Mass is assumed to be uniformly distributed within each tetrahedron and the tessellation remain
intact at all times. The tetrahedra of the tessellation change their shapes and volumes, the latter are used for
computing the densities of the tetrahedra. Despite the complicated deformations experienced by the three-dimensional
sub-manifold tessellated by the tetrahedra, it remains continuous. Projected on three-dimensional configuration space, the tetrahedra
may form complicated structures. The number of streams at a chosen point {\bf x} is simply the number of tetrahedra that contain the point. The diagnostic points are computationally convenient to choose on a regular grid which can be significantly finer than the original grid in Lagrangian space. The ratio of separation of particles on the initial unperturbed grid to the separation distance of points in the diagnostic grid  $l_{\rm part}/l_{\rm dg} $ will be referred to as the refinement factor in the rest of the paper.

Number of streams are odd-valued in the entire configuration space, except in a set of points of measure zero where caustics are formed.  
 A single-stream flow implies that the tetrahedra do not overlap in the corresponding region and thus defined as a void region.
 The web is defined as a set of non-void regions, i.e. the set of regions where the number of stream is equal to or more than three.
 The level of non-linearity in the web  can be quantitatively characterized by using `number of streams' as a parameter. As shown in  \citep{Shandarin_etal:12} there is no simple local relation between the number of streams and density, however the both
 fields are obviously correlated.

\section{Global statistics of Cosmic web}
\label{sec:global}

The 3-dimensional multi-stream field for the entire simulation box exhibits cosmic web structure with void, walls, filaments and haloes. We propose the number of streams, `$n_{str}$' as a parameter for characterizing and distinguishing structures in the universe. This is different from \citet{Flack_Neyrinck:14}, 
where the authors have identified voids, walls, filaments and haloes by particles which have undergone any number of flip-flops along 0, 1, 2 or 3 axes respectively. 
Their description of voids is close to ours except that some particles that have  experienced no flip-flops might be in the region
of multi stream flow formed by other particles.
Thus we expect that the mass fraction in voids defined as the regions with $n_{str} = 1$ is somewhat lower than that defined as the particles with 
flip-flops $= 0$ only at the final state. This is because some particles may have already fallen in the web but  have not experienced flip flop yet and 
some particles that have experienced an even number of flip flops may come back to the original order. However the above arguments are valid only if the thickness of the web is the same in the both approaches. As we discuss below the thickness of the web in our analysis is about (100\% - 84\%)/(100\%-93\%)=2.3 times thinner than in  the analysis by \citet{Flack_Neyrinck:14} (See Table \ref{tab:Compare} for details). However  \citet{Flack_Neyrinck:14} discussed these effects and claimed that they were small.

Whereas for non-linear structures, our parameter space has more freedom in terms of number of streams. 
Similar to the density threshold, the number of streams - used as a local parameter - cannot distinguish unambiguously whether a point is in a wall, 
filament or halo. Only some parts of walls where there are only three streams can be identified locally without confusion. This is because the formation of a filament
requires at least five streams. A flip-flop along one axis  would produce a three-stream region which may be only a pancake. 
 Therefore  another flip-flop along the other axis in one of the streams
from previous stage is required to transform it into a three-stream flow. Thus the total becomes five. However, if the second flip-flop happens along the same axis 
the resultant structure will remain a wall. Therefore some points in the five-stream flows can be within walls while the other in filaments. The present simulations have no information about the evolution of the flip-flop field
therefore we rely on visual impressions initially to understand the transformation of walls into filaments and parts of filaments into  haloes. By inspection, we have identified all the regions with three streams as walls. Unfortunately, walls are difficult to display on paper since they essentially block the view in two-dimensional projection. Nevertheless, we have demonstrated and analysed walls on a smaller Eulerian box around haloes in Section \ref{sec:local} using a simple and reasonably effective approximation.


For a multi-stream field calculated on a simulation box of size $100 h^{-1}$ Mpc and $128^3$ particles, it is visually observed that with the increase of $n_{str}$ from 3 to 15,  the corresponding occupied regions increasingly belong to filamentary structure rather than the membrane like walls, until at the level $n_{str}  \gtrsim 17$ we observe that the number of wall points become negligible.

\begin{figure}
 \centering\includegraphics[width=8.5cm]{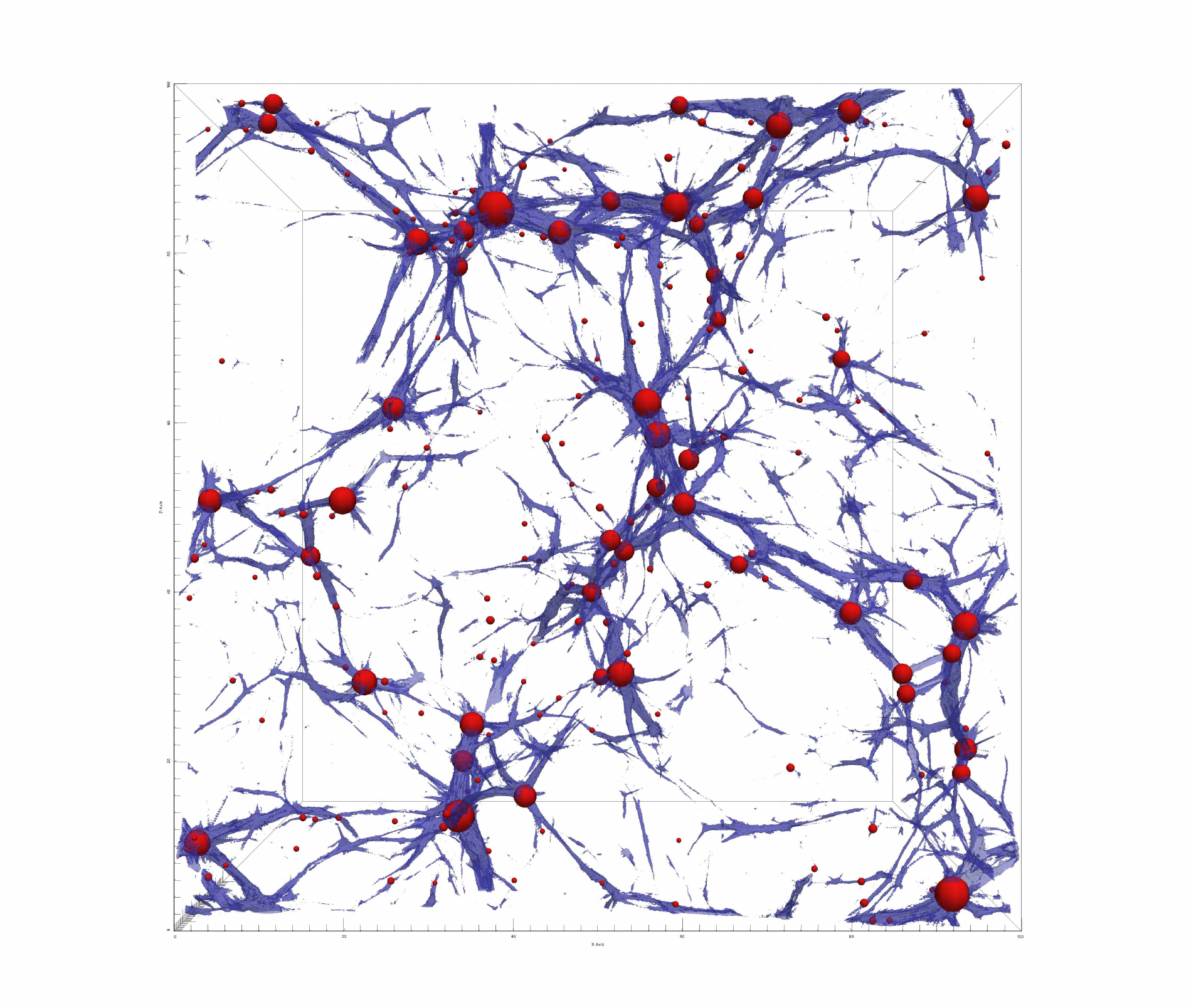} 
\caption{The cosmic web structure in a slice of  $30 h^{-1} \text{Mpc} \times 100 h^{-1} \text{Mpc} \times 100 h^{-1} \text{Mpc}$ in a simulation box of size $100 h^{-1}$ Mpc and $128^3$ particles. Regions with 17 or more streams (blue) form a filamentary structure. The haloes determined by FOF (red) are predominantly embedded in the intersections of the filaments.}
\label{fig:full}
\end{figure}

The filamentary structure of regions with 17 or more streams (denoted as 17+ ) is shown for a slice of simulation box of size $100 h^{-1}$  Mpc and $128^3$  particles in Fig. \ref{fig:full}. It has to be noted that all the regions with 17+ streams are regions with 3+ streams. Thus, the filaments are just interior
parts  of walls with higher $n_{str}$. These are visually observed mostly at the intersections of walls. Further, at the intersections of multiple filaments, there are regions with locally maximum number of streams, signifying the most dense regions in the simulations i.e. the dark matter haloes as Fig. \ref{fig:4view}
illustrates.
By superimposing the positions from the FOF-halo catalogue, it is visually confirmed that the FOF haloes reasonably coincide with these high-streaming intersections, as Fig. \ref{fig:full} illustrates.

\subsection{Volume and mass fractions}

\begin{figure}
\begin{minipage}[t]{.99\linewidth}
  \centering\includegraphics[width=8.cm]{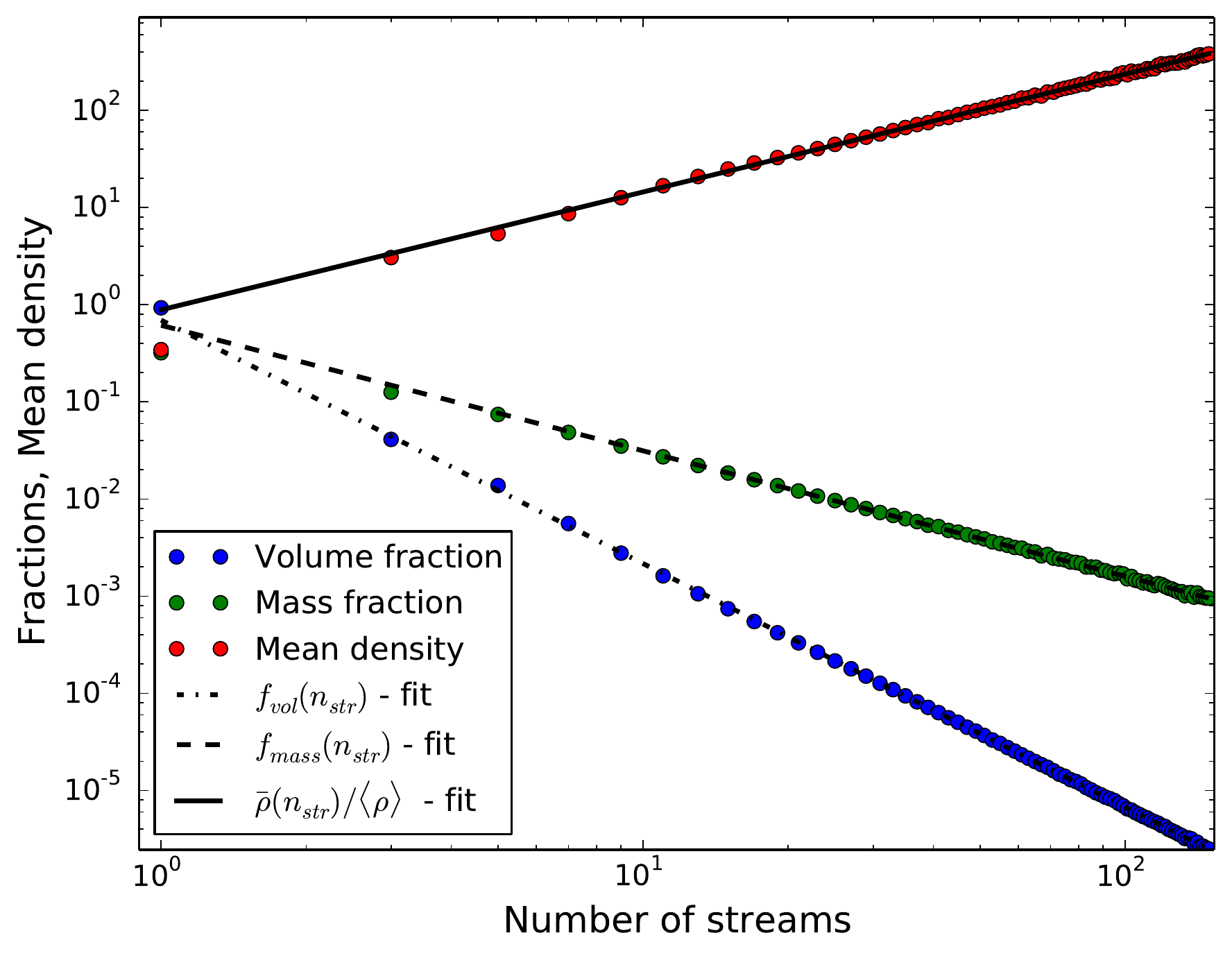} 
\end{minipage}\hfill
\caption{Volume and mass fraction of each stream, mean density of each stream in a box of size 100 $h^{-1}$ Mpc and  $128^3$ particles. Exact values of fractions, density and their curve fit for the range $n_{str} \ge 5 $ from Eq.~\ref{eq:fr_vol}, Eq.~\ref{eq:fr_mass} and Eq.~\ref{eq:den} are shown. Multi-streams are calculated with refinement factor of 8. The void ($n_{str} = 1$) occupies 93$\%$ of the volume and 55$\%$ of the mass. }
\label{fig:m_v_fr}
\end{figure}

The single-stream flow, which corresponds to the void, occupies majority of volume of the simulation box (Fig. \ref{fig:m_v_fr}). As mentioned in Section \ref{sec:global}, higher multi-streaming flow regions are nested inside the lower streaming regions. Thus the volume occupied by higher number of streams monotonically decreases with the number of streams. This relation is approximately found to be a power law. For the box of size $L =100 ~h^{-1}$ Mpc and $N = 128^3$ particles ($L/N = 0.78 ~h^{-1}$ Mpc), the volume fraction corresponding to each value of number of streams, $ f_{vol}(n_{str})$ in the multi-stream field calculated with refinement factor of 8 (i.e. the multi-stream field
was computed on 1024$^3$ grid as described in \cite{Shandarin_etal:12}) is 
\begin{equation}
\label{eq:fr_vol}
f_{vol}(n_{str}) =  0.69 n_{str}^{-2.5}
\end{equation}

This is a good fit for the range of number of streams $n_{str} \ge 5 $. In multi-stream field for the simulation box mentioned above, about 93$\%$ of the volume is occupied by 1-stream. With an increase in $n_{str}$, the corresponding volume fraction reduces. Physically, however, the number of streams reflect the advancement of non-linearity. Hence the higher $n_{str}$ regions are typically regions with higher densities. In effect, the mass fraction can also be approximated by a decreasing power law function of $n_{str}$,  
\begin{equation}
\label{eq:fr_mass}
f_{mass}(n_{str}) = 0.61 n_{str}^{-1.3}
\end{equation}
This is also a good fit for the range of number of streams $n_{str} \ge 5 $. For the same range of number of streams, the mean density in the regions with particular number of streams, given by the  ratio of corresponding mass and volume fractions, increases as expected.
\begin{equation}
\label{eq:den}
{\bar{\rho}(n_{str}) \over \langle{\rho}\rangle} = 0.89  n_{str}^{1.2} ,
\end{equation}
where $\langle\rho\rangle$ is the mean density of the universe.
This also quantifies our previous claim that very high multi-streams correspond to the most dense areas in the Universe, i.e. the condensed haloes. The common over-density threshold of 200 using virial equilibrium corresponds to roughly 90 streams in  Fig. ~\ref{fig:m_v_fr} and Eq. ~\ref{eq:den}.

\begin{figure}
\begin{minipage}[t]{.99\linewidth}
 \centering\includegraphics[width=8.cm]{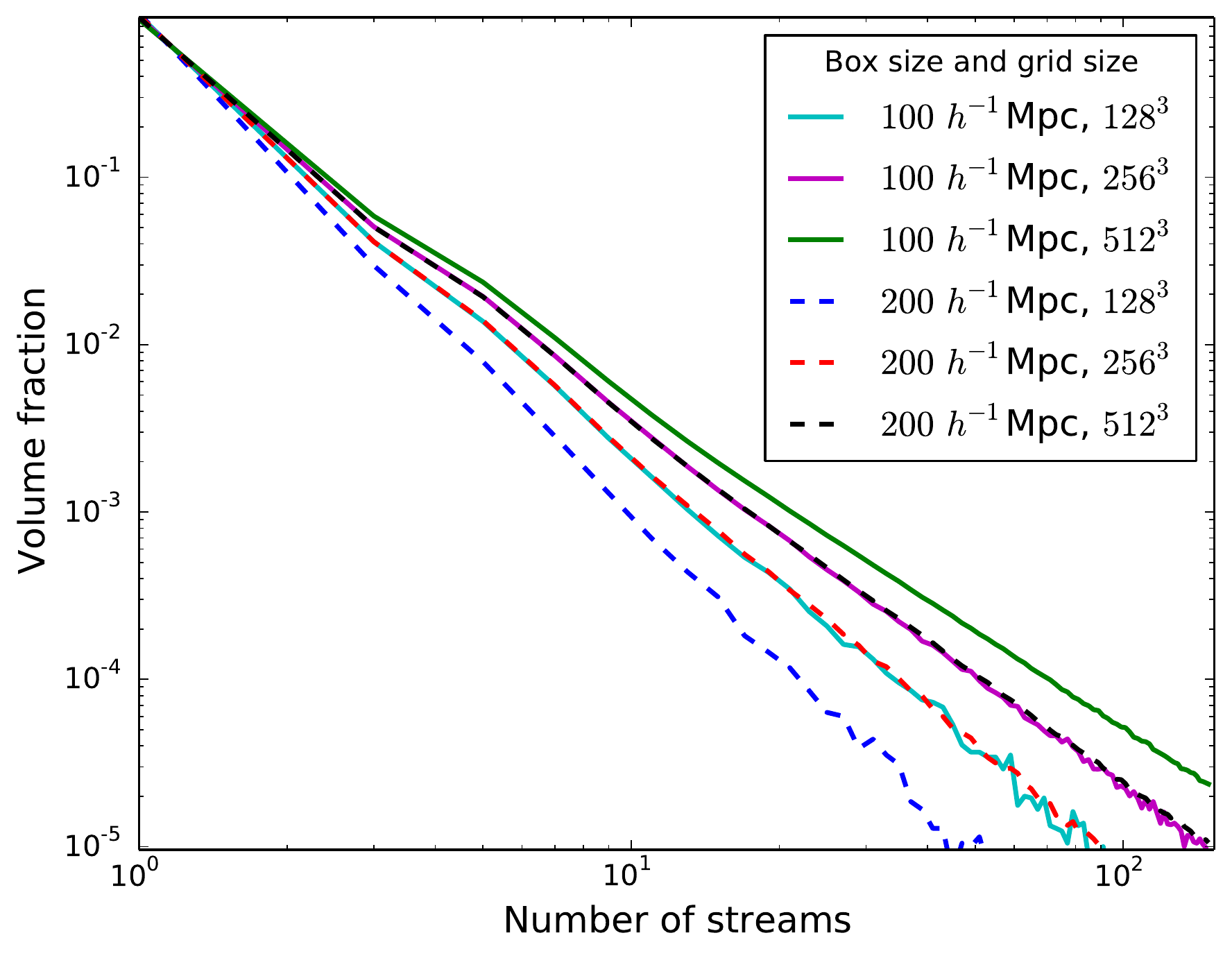} 
  \centering\includegraphics[width=8.cm]{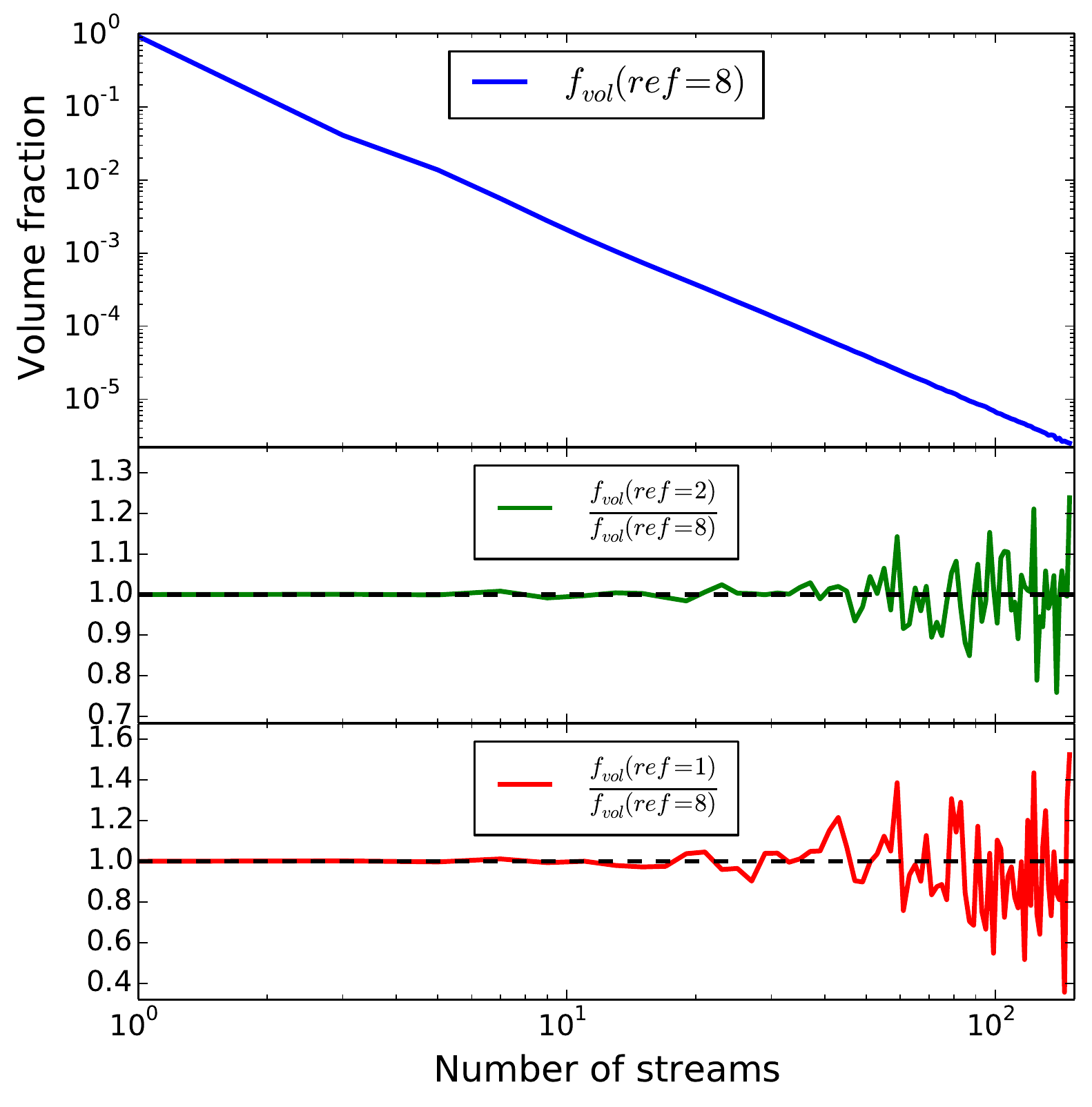}
\end{minipage}\hfill
\caption{Top: Volume fraction distribution of streams in 6 simulation boxes of size 100 $h^{-1}$ Mpc, 200 $h^{-1}$ Mpc 
and $128^{3}$, $256^{3}$, $512^{3}$ grids (with refinement factor of 1). Volume fractions are similar for simulation boxes with same inter-particle resolution. Slopes of the curve fits are shown in Table ~\ref{tab:Compare_Slopes}. Bottom: Volume fraction distribution for different refinement factors for 100 $h^{-1}$ Mpc, $128^{3}$ box. A considerably smoother volume fraction distribution is obtained at high number of streams in multi-stream fields with high refinement factor. }
\label{fig:Vfr_all}
\end{figure}

Comparing the volume fractions of various simulation boxes in Fig. ~\ref{fig:Vfr_all} and corresponding power law dependences in Table ~\ref{tab:Compare_Slopes} (also, specifically for the volume fraction of voids in Table ~\ref{tab:Compare_LN}), we find that the profile is similar for boxes with
same inter-particle resolution; i.e., equal box length to grid size ratio( For e.g., $L/N = 0.78$ $h^{-1}$ Mpc for the simulation box of 100 $h^{-1}$ Mpc - $128^3$ particles and 200 $h^{-1}$ Mpc - $256^{3}$ particles). The box with minimum inter-particle resolution in the data, hence the best raw resolution ( $L/N = 0.19$ $h^{-1}$ Mpc for 100 $h^{-1}$ Mpc, $512^3$ particles), has higher volume fraction for each multi-stream compared to lower resolution boxes. Additionally, it has a more non-linear stage advanced over time resulting from the initial small scale perturbations. The advancement of  non-linearity manifests itself in higher number of streams. Box with the least raw inter-particle resolution ($L/N = 1.56 $  $h^{-1}$ Mpc for 200 $h^{-1}$ Mpc, $128^3$ particles), occupies lower volumes than other boxes for each $n_{str}$. It is also prone to noise at very high streaming regions.

One of the advantages of using tessellation is the freedom to compute densities at very high resolutions (\cite{Abel_etal:12}, \cite{Shandarin_etal:12}). We remind that the parameter `refinement factor' denotes the ratio of separation of the particles to the separation distance of points in the diagnostic grid as defined in Sec. 3. High refinement factors are extensively used in understanding stream behaviour not only in the halo environment, but inside the halo too (Section ~\ref{sec:local}). The volume fractions of resulting number of streams are
similar for all refinement factors as
shown in bottom of the Fig. ~\ref{fig:Vfr_all} and in Table ~\ref{tab:Compare_ref}. 
Multi-stream fields calculated on low refinement factors are more noisy at high number of streams.

\begin{figure}
\begin{minipage}[t]{.99\linewidth}

  \centering\includegraphics[width=8.cm]{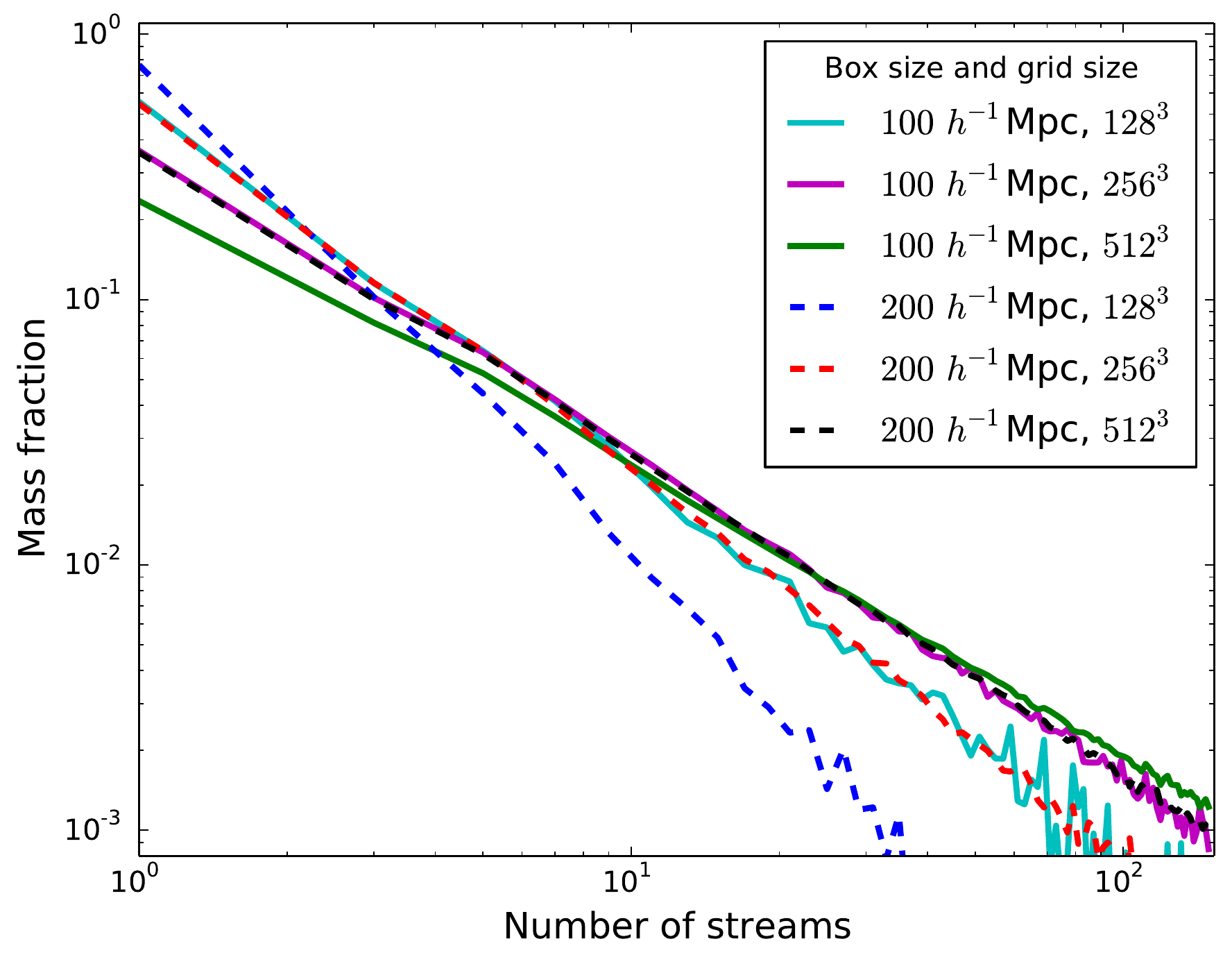} 
  \centering\includegraphics[width=8.cm]{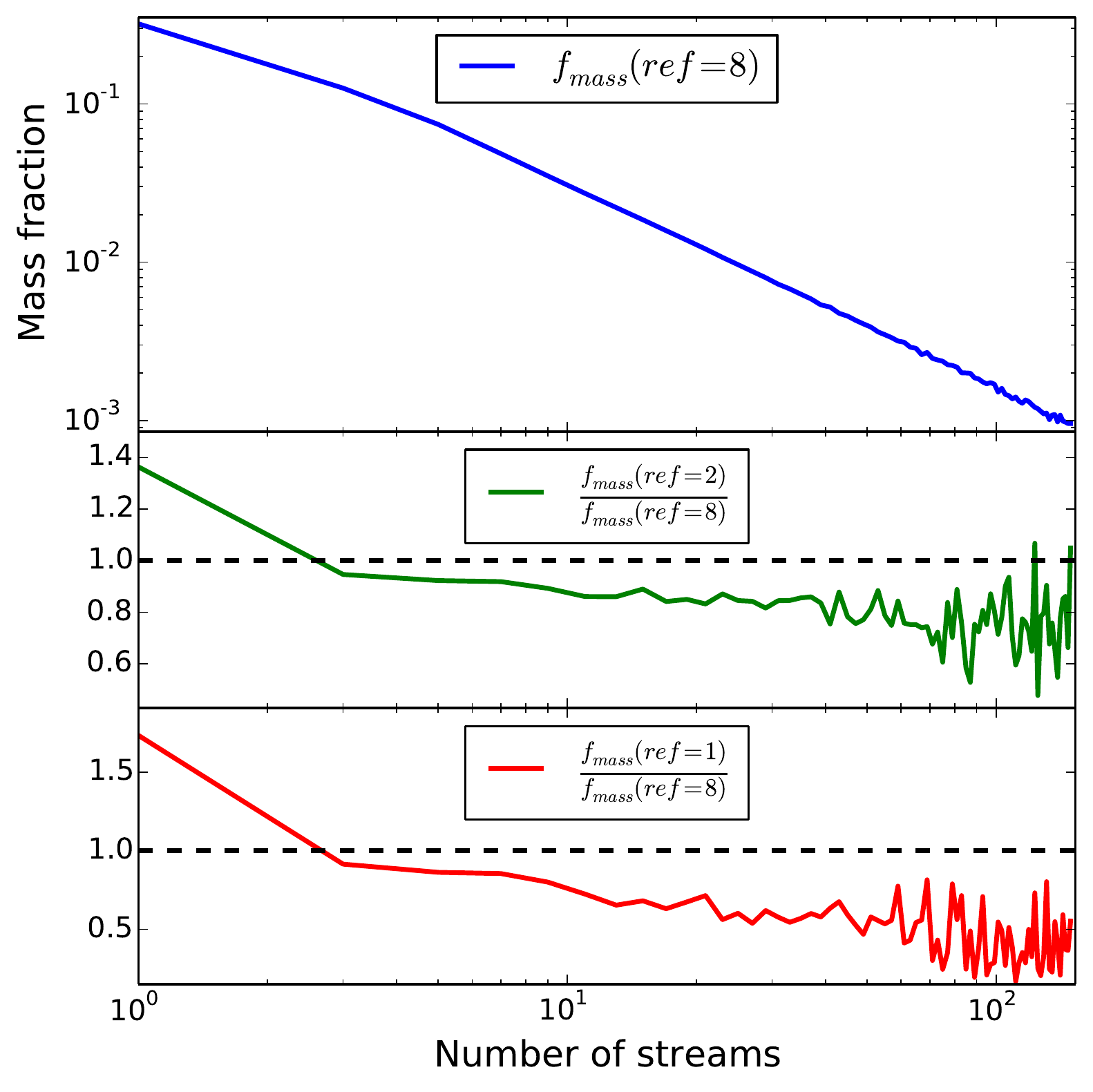}
\end{minipage}\hfill
\caption{Top: Mass fraction distribution of streams in 6 simulation boxes of size 100 $h^{-1}$ Mpc, 200 $h^{-1}$ Mpc 
and $128^{3}$, $256^{3}$, $512^{3}$ grids ( with refinement factor of 1). Mass fractions are similar for simulation boxes with same inter-particle resolution. Slopes of the curve fits are shown in Table ~\ref{tab:Compare_Slopes}. Bottom: Mass fraction distribution for different refinement factors for 100 $h^{-1}$ Mpc, $128^{3}$ box. Single-streaming void regions have more fraction of mass particles in multi-stream fields calculated at low refinement factors of 1 and 2. This effect is minimized when calculation is done at better refinement. In addition, mass fraction distribution is less noisy.}
\label{fig:Mfr_all}
\end{figure}

With same refinement factors, the mass fractions exhibit similar pattern for same $L/N$ as well (Fig.~\ref{fig:Mfr_all}). The simulation box with highest inter-particle distance (thus least mass resolution) has more mass particles in single streaming region, as tabulated in Table~\ref{tab:Compare_LN}, but decreases steeply thereafter (Table ~\ref{tab:Compare_Slopes}). Unlike the volume fraction, the behaviour of mass fraction has a systematic variation across different refinement factors. The mass fractions given in Table ~\ref{tab:Compare_ref} show that the single-streaming regions in the multi-stream fields with refinement factors 1 and 2 have higher mass fraction than in the fields with refinement factor of 8. Decreasing the resolution from refinement factor of 8 to 1 effectively introduces smoothing of the structure. This results in growth of mass fraction in voids ($n_{str} = 1$) and decreasing it in the web ($n_{str} > 1$). The multi-stream field is more robust as one can see in Fig.~\ref{fig:Vfr_all}. and \ref{fig:Mfr_all}. In addition, the less refined multi-stream grids are prone to noise at high $n_{str}$ as usual.

\begin{table}
  \caption{Comparison of the approximate power law dependences of curve fits in Fig. \ref{fig:Vfr_all} and Fig. \ref{fig:Mfr_all}. Power law relations for volume fraction $f_{vol}(n_{str})$ and mass fractions $f_{mass} (n_{str}) $ as a function of number of streams at $n_{str} \geq 5$ are shown (amplitudes are not shown). The boxes of size 100 $h^{-1}$ Mpc on  $128^{3}$ grids and, 200 $h^{-1}$ Mpc on  $256^{3}$ grids have same  $L/N = 0.78 h^{-1}$ Mpc. Similarly, $L/N = 0.39 h^{-1}$ Mpc for boxes of size 100 $h^{-1}$ Mpc on  $256^{3}$ grids and, 200 $h^{-1}$ Mpc on  $512^{3}$ grids.}
\begin{tabular}{|l|r|r|r|r|}
\hline
$L/N$                & $0.19$ & $0.39$& $0.78$ & $1.56$  \\ \hline
$f_{vol} (n_{str}) $ Vs. $n_{str}$                 & $-2.1$   & $-2.3$    & $-2.5$      & $-2.9$  \\ \hline
$f_{mass} (n_{str}) $   Vs. $n_{str}$            & $-1.1$   & $-1.2$   & $-1.4$       & $-2.0$     \\ \hline
\end{tabular}
 \label{tab:Compare_Slopes}
\end{table}

 \begin{table}
  \caption{Comparison of the volume and mass fractions of the void ($n_{str} = 1$) regions of the cosmic web for various simulation boxes at refinement factor of 1. Mean density is the ratio of mass fraction to the volume fraction. It is given in units of the mean density of the universe. The boxes of size 100 $h^{-1}$ Mpc on  $128^{3}$ grids and, 200 $h^{-1}$ Mpc on  $256^{3}$ grids have same  $L/N = 0.78 h^{-1}$ Mpc. Similarly, $L/N = 0.39 h^{-1}$ Mpc for boxes of size 100 $h^{-1}$ Mpc on  $256^{3}$ grids and, 200 $h^{-1}$ Mpc on  $512^{3}$ grids.}
\begin{tabular}{|l|r|r|r|r|}
\hline
$L/N$                & $0.19$ & $0.39$& $0.78$ & $1.56$  \\ \hline
Volume Fraction (\%)                 & $88$   & $90$    & $93$      & $96$  \\ \hline
Mass Fraction   (\%)               & $24$   & $36$   & $55$       & $77$     \\ \hline
Mean density                    & $0.27$ & $0.40$    & $0.59$        & $0.80$   \\ \hline
\end{tabular}
 \label{tab:Compare_LN}
\end{table}

 \begin{table}
  \caption{Comparison of the volume and mass fractions of the void ($n_{str} = 1$) regions of the cosmic web for a simulation box at different refinement factors. Mean density is the ratio of mass fraction to the volume fraction. All the multi-streams for simulation box of length 100 $h^{-1}$ Mpc on raw resolution of $128^{3}$ grids ($ L/N = 0.78 h^{-1}$) }
\begin{tabular}{|l|r|r|r|}
\hline
Refinement factor               & $1$ & $2$& $8$   \\ \hline
Volume Fraction (\%)                 & $93$   & $93$    & $93$       \\ \hline
Mass Fraction   (\%)               & $55$   & $44$   & $32$           \\ \hline
Mean density                    & $0.59$ & $0.47$    & $0.35$        \\ \hline
\end{tabular}

 \label{tab:Compare_ref}
\end{table}

\section{Stream environment around haloes}
\label{sec:local}

Multi-stream field can be easily computed for a small Eulerian box with higher refinement factor. This can be utilized to analyse the phase-space behaviour inside and around haloes. In this section, 
we have used the halo coordinates identified by the FOF method, and selected Eulerian boxes around it. A reasonable  correspondence between FOF halo centres and local maxima of multi-stream field is visually examined in Fig. \ref{fig:full}. 

\begin{figure}
\centering
\begin{minipage}[c]{.45\linewidth}
\includegraphics[width=4.5cm]{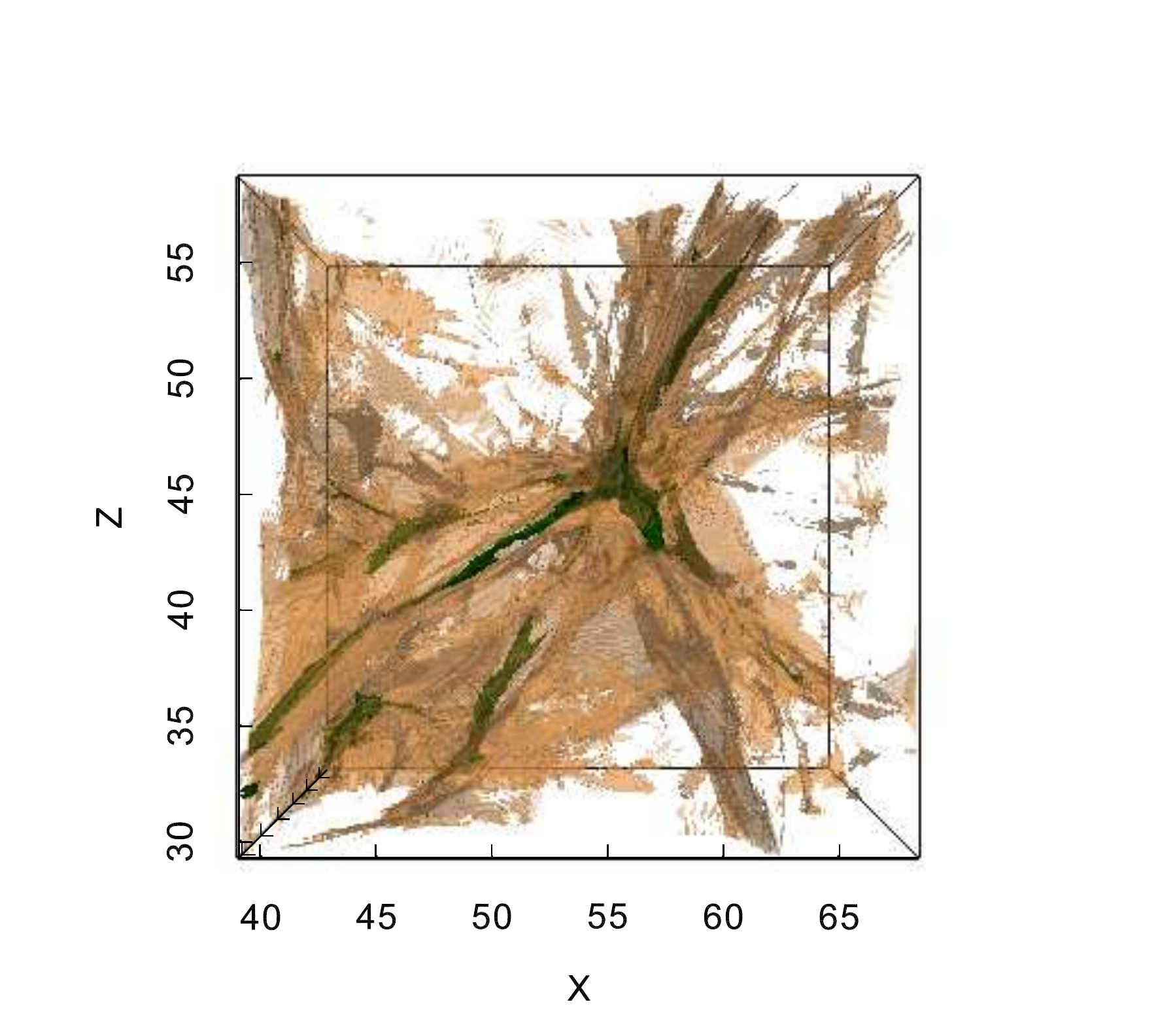} 
\end{minipage}
\begin{minipage}[c]{0.45\linewidth}
\includegraphics[width=4.5cm]{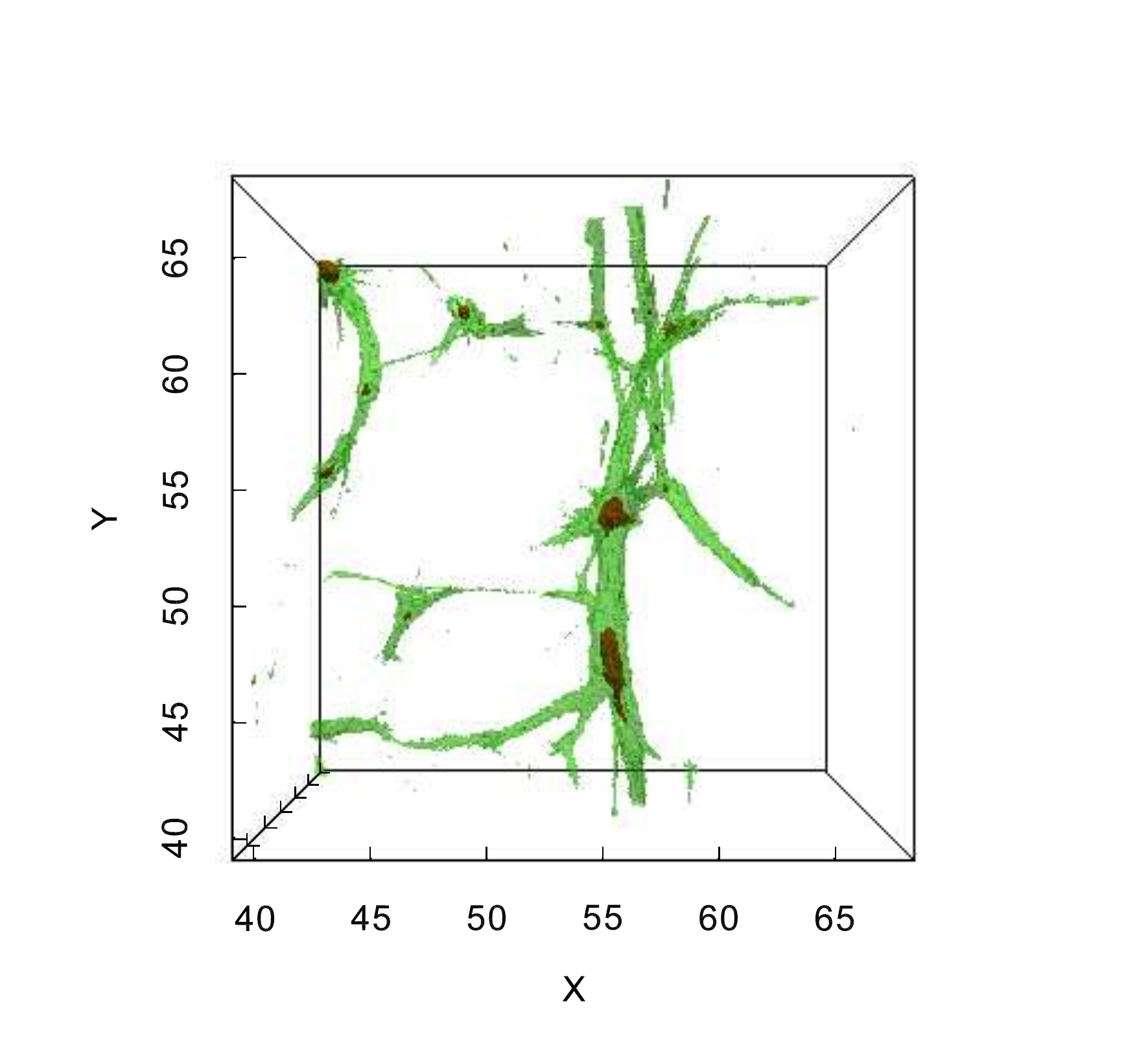} 
\end{minipage} 
\begin{minipage}[c]{0.45\linewidth}
\includegraphics[width=4.5cm]{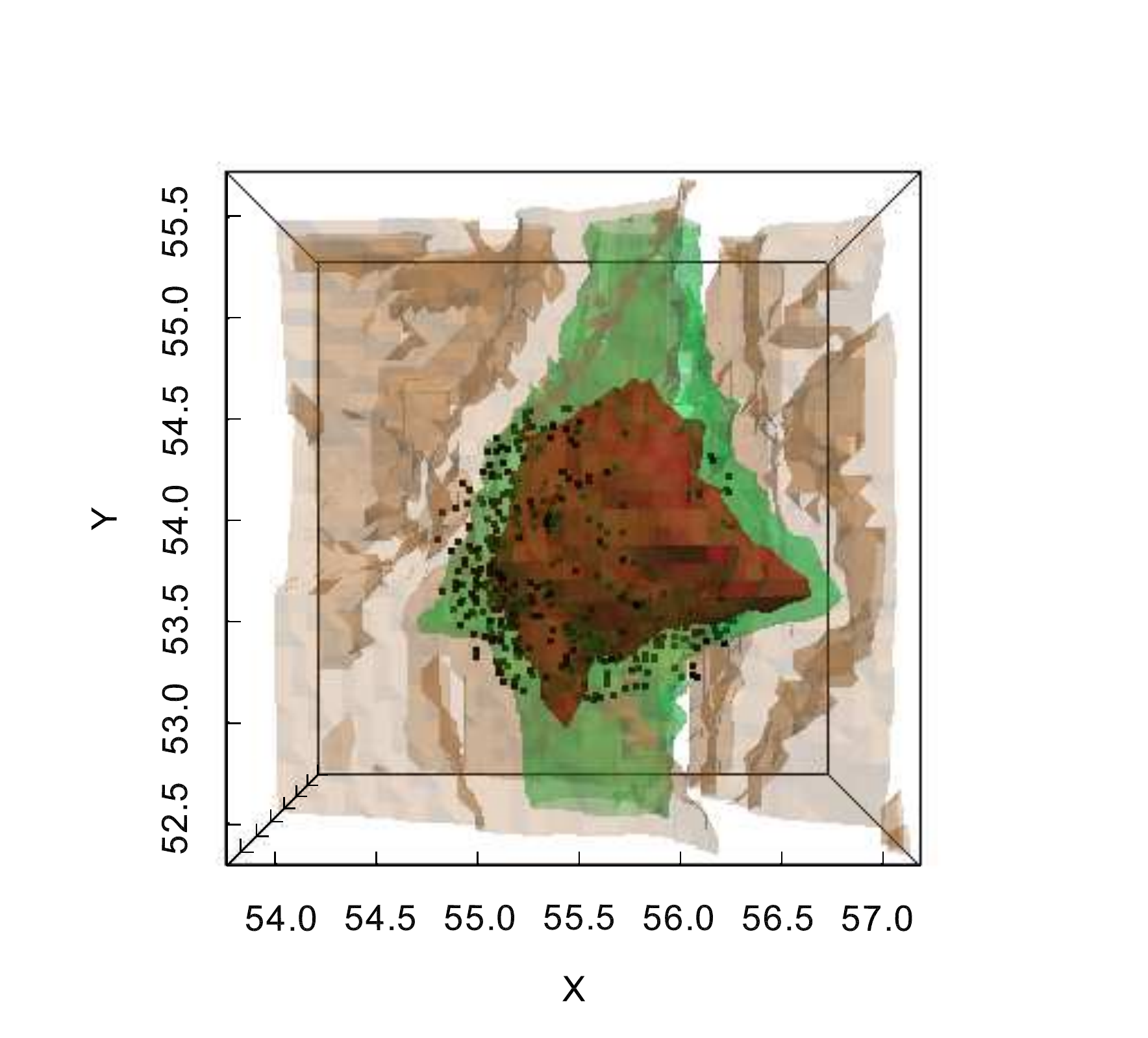} 
\end{minipage}
\begin{minipage}[c]{0.45\linewidth}
\includegraphics[width=4.5cm]{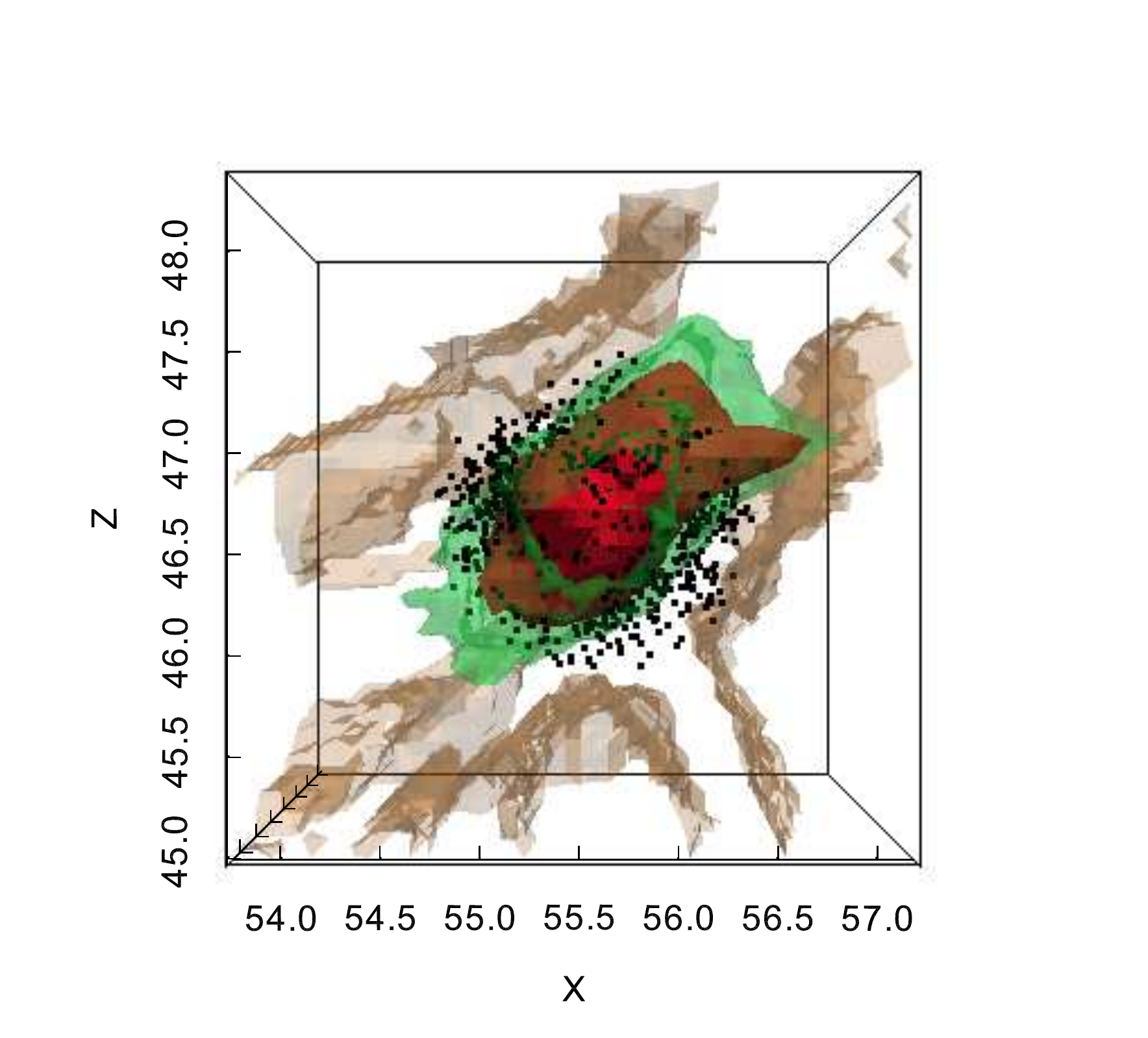} 
\end{minipage} 

\caption{Multi-stream flow regions in a small box of the simulation. Top left: regions with more than 3-stream flow are identified as walls (brown). Intersection of multiple walls have higher $n_{str}$ regions (green, 17+ streams). Single-streaming voids (white) occupy large volume and are very close to the filaments in some directions. Top right: 17+ streams (green) form filamental structures with nodes at the intersections (red, 90+ streams)
Bottom left and right:  Closer look at the highly non-linear region reveals that a filament is sandwiched between the walls (brown). The 90+ stream region (red) forms a compact structure and is entirely contained within the filament. The black dots show the particles around the FOF halo within linking length of 0.2.}
\label{fig:4view}
\end{figure}

Since each multi-stream region is surrounded by lower number of streams, the walls sandwich filaments within themselves (Fig. \ref{fig:4view}). The filaments are embedded with haloes at various intersections. These high-streaming haloes are completely covered by relatively low-stream filaments and hence surrounded by walls too. This result differs considerably from the several void finder methods, which find existence of haloes within  void regions (See \citealt{Colberg_etal:08} and references therein). By our classification, we distinguish configuration space of the simulation box as void and non-void or web regions. Further, we have made an attempt to classify the web into walls, filaments and haloes based on multi-stream thresholds. This classification based on number of stream threshold provides only a very crude description of visual impression from the richness, complexity and fundamentally multi-scale character of the web. The heuristic numbers we use in this paper are by no means universal, but may provide limited use in the discussion of these particular simulations.

Visual inspection of Fig. \ref{fig:4view} reveals that the multi-stream environment of a halo is a highly intricate. Though the haloes are surrounded by filaments and walls, it can be surprisingly close to the voids in particular directions. Filaments 
defined by the multi-steam field are  quite elongated, but the cross-sections are not circular or elliptical, and moreover, they branch-out and intersect at several regions. Finally, the  haloes defined by contours of the multi-steam field look neither spherical nor ellipsoidal. 
We use a simple geometrical technique of projecting the number of streams onto a diagnostic spherical surface around 
a haloes to visualize and quantify their environments.

\subsection{Technique}

Motivated by the complicated morphology of multi-stream field around a dark matter halo, we devised an empirical 
statistical tool to quantify the multi-stream environment of the FOF haloes. The method is geometrical and non-local.
We  randomly select a large number of points on a diagnostic spherical surface centred at the FOF centre of the halo 
and compute the number of streams  at these points. By increasing the radius of the sphere from inside of the halo to several times the halo radii, we estimate  the fractions of the area on the diagnostic spherical surface occupied by the regions with different numbers of streams:
3+, 5+, ..., where $n+$ corresponds to $n$ or higher number of streams.

\begin{figure}
\begin{minipage}[t]{.99\linewidth}
 \centering\includegraphics[width=6.cm]{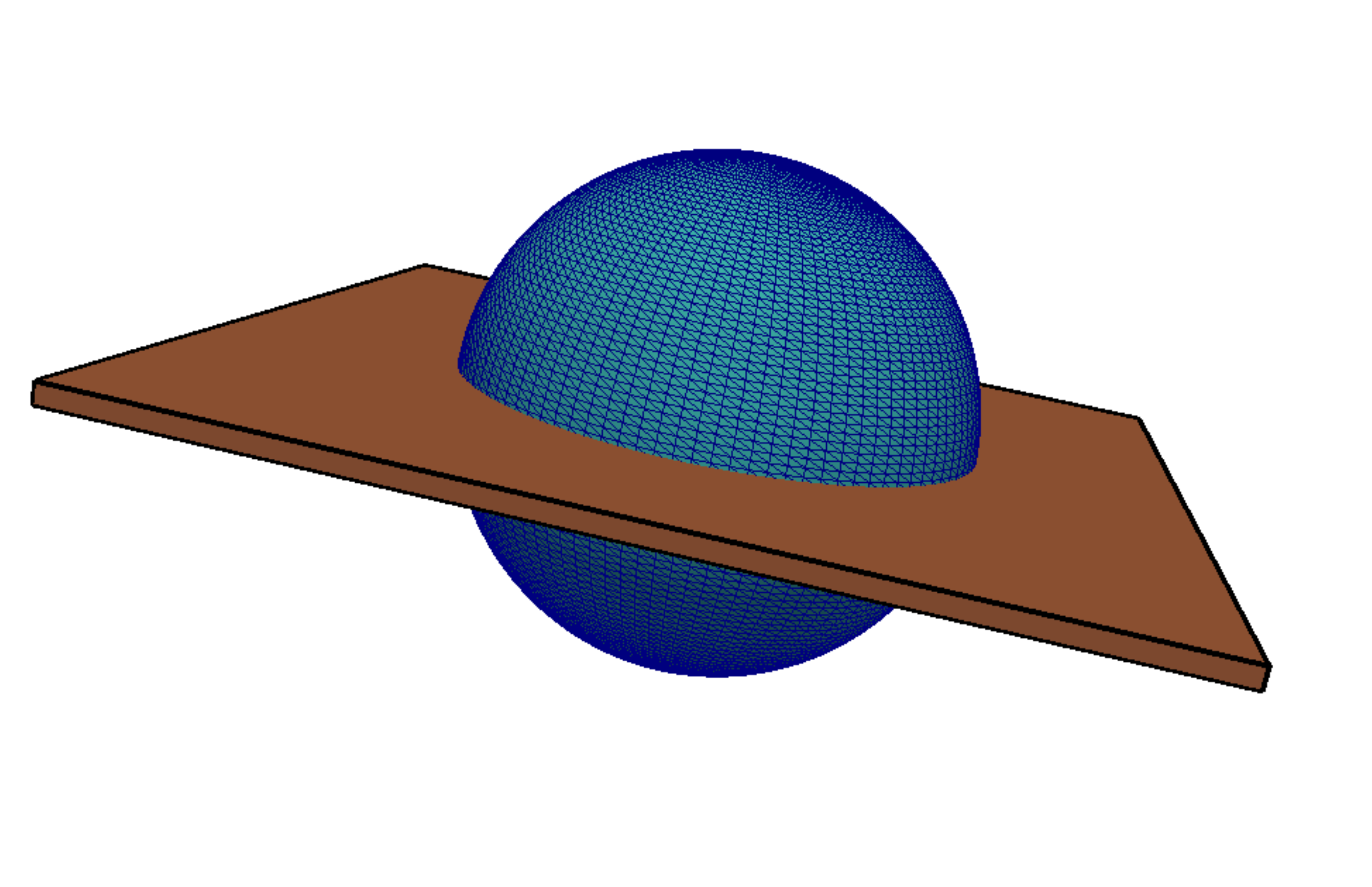} 
  \centering\includegraphics[width=6.cm]{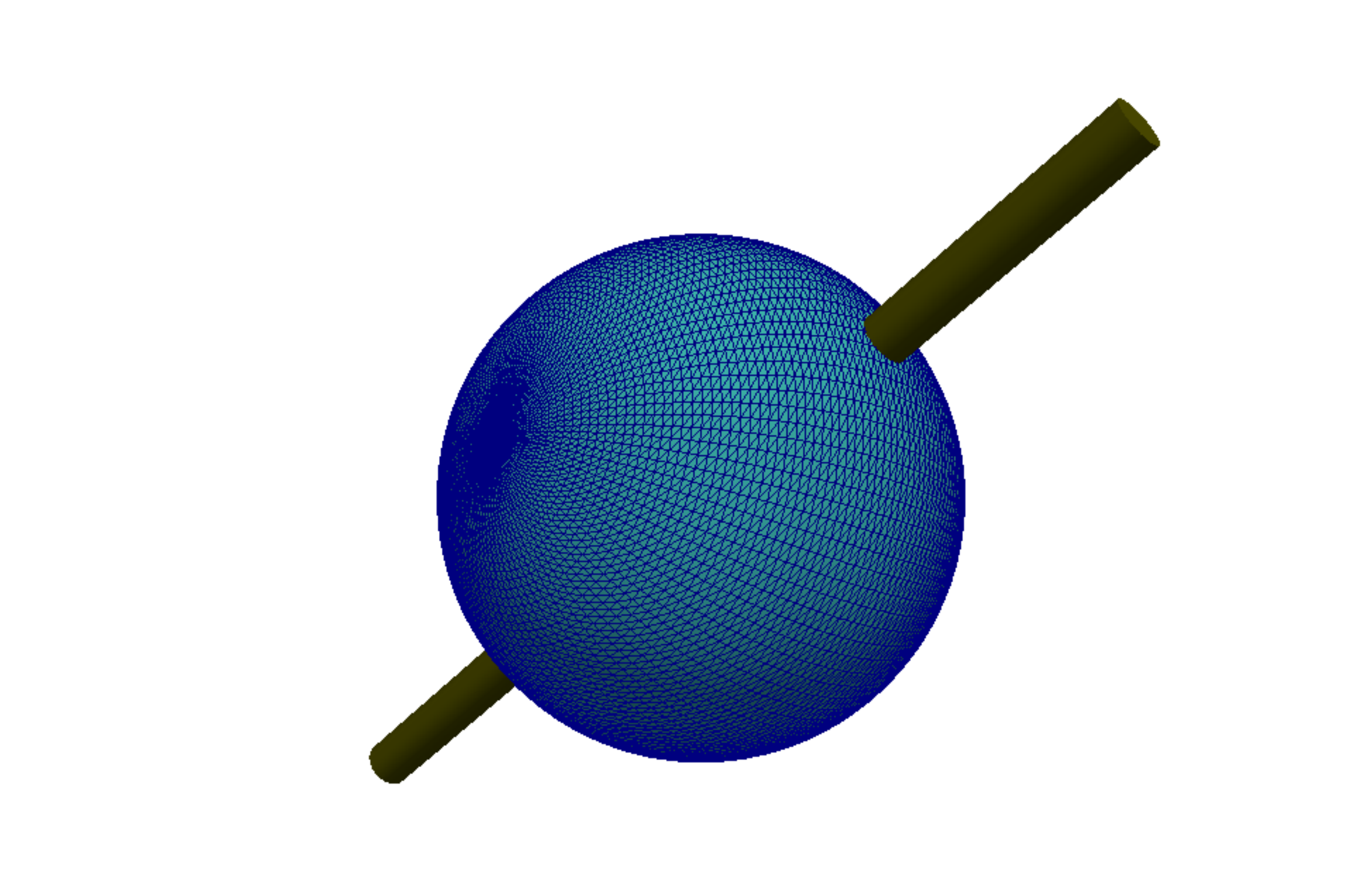}
\end{minipage}\hfill
\caption{Modelling a wall and a filament. A diagnostic spherical surface is intersected by a cylinder and plane.}
\label{fig:model}
\end{figure}

The geometry of a filament can be crudely approximated by a cylinder and that of a wall by a sheet with a small constant thickness 'd'. Upon intersecting with the spherical surface, these geometries occupy certain cross-sectional area, $Area_{c/s}$, on the sphere (See Fig. ~\ref{fig:model}). The ratio of this area to the surface area of the sphere  is given by,
\begin{subequations} \label{eq:model}
\begin{align}
 f_{wall} (r) &= \frac{Area_{c/s}}{4\upi r^2} = \frac{2\upi r d}{4\upi r^2} \propto r^{-1} \\
 f_{fil} (r) &= \frac{Area_{c/s}}{4\upi r^2} = \frac{\text{const.}}{4\upi r^2} \propto r^{-2}
\end{align} 
\end{subequations}
The fractions of points on the surface of the sphere by multiple number of intersecting sheet-like walls or cylindrical filaments also scale proportional to $ r^{-1}$ and $ r^{-2}$ respectively. 

For the diagnostic spheres of different radii, the  scaling of multi-streams at the intersections is calculated. By checking the 
variation in the fraction of area occupied, we associate the number of streams with wall or halo. 

Each of the Mollweide projections in Fig. ~\ref{fig:3} - ~\ref{fig:4322} shows projection of the multi-stream field on to the 
spherical surface, and provide useful insight into the multi-stream structure around a halo. 
In a Mollweide projections, each filament stemmed from the halo looks as a compact patch. If the physical area of the cross 
section of the filament remains approximately constant, then the size of the patch on the Mollweide projection would decrease with the increasing radius of the diagnostic sphere. The cross section of a wall with the diagnostic sphere has a well known
`S'-shape (similar to the ecliptic plane in the galactic coordinates) and the width decreases with the growth of the diagnostic sphere. Both patterns are clearly seen in  Fig. ~\ref{fig:3} - ~\ref{fig:4322}.

\subsection{Voids, filaments and walls around haloes}

From the technique described  above, we arrive at quantitative thresholds for the different components of the web i.e., all regions where
$n_{str} \ge 3$. We stress that this method is only a practical tool in arriving at heuristic thresholds of cosmic web structures. The analysis done here are for the simulation box of 100 $h^{-1}$ Mpc, 128$^3$ particles, with refinement factor of 8.

\begin{figure}
\begin{minipage}[t]{.99\linewidth}
  \centering\includegraphics[width=8.cm]{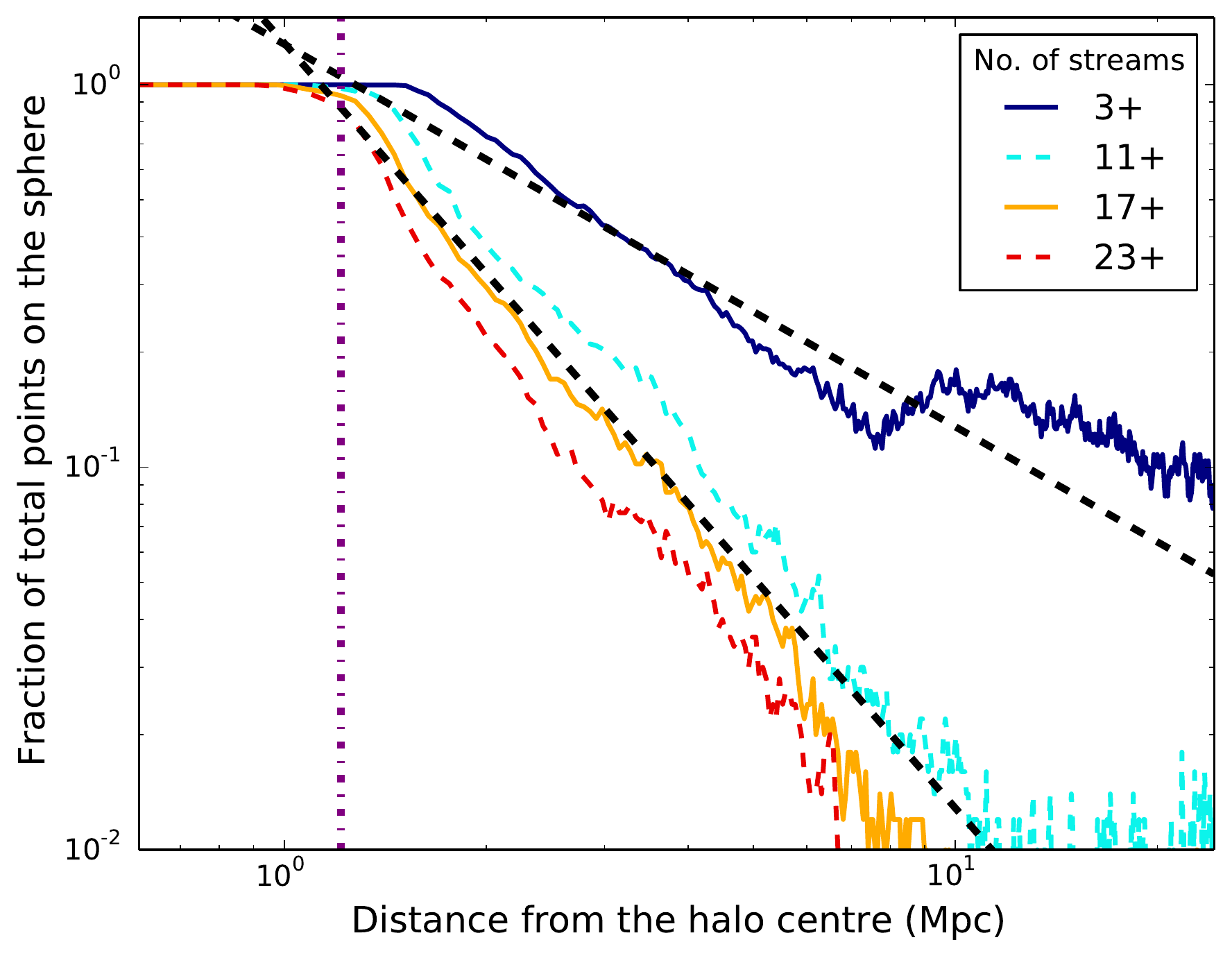}
\includegraphics[width=8.cm]{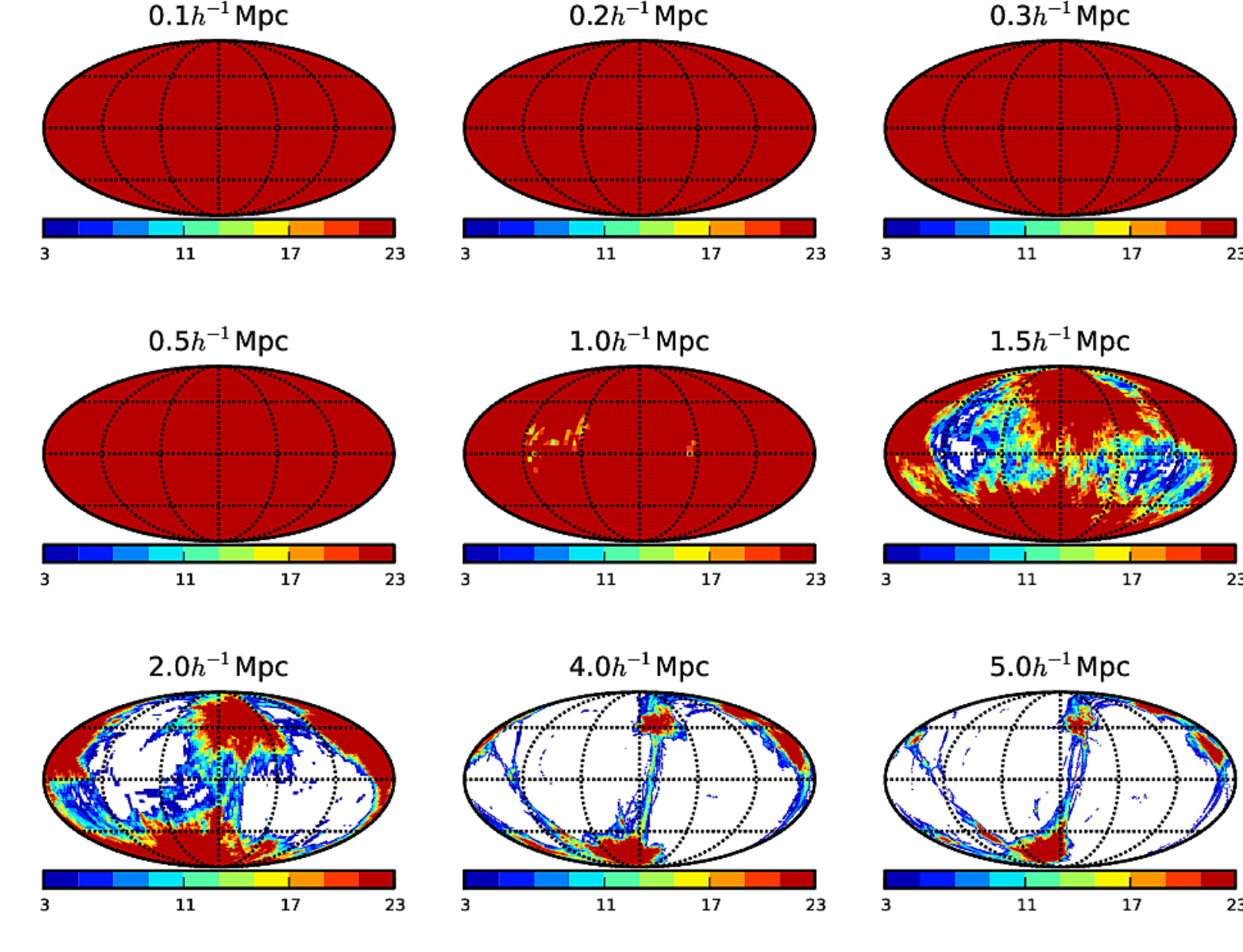} 
\end{minipage}\hfill
\caption{ Large halo of mass $3.7 \times 10^{14}M_{\odot} $  and FOF radius 1.2 $h^{-1}$ Mpc (dotted-violet line in the top figure). Top: Fractional distribution of streams on the surface of spheres of increasing radii. Dashed-black lines are for $r^{-1}$ and $r^{-2}$ scaling. 3+ streams are closest to $r^{-1}$ and 17+ scales close to $r^{-2}$. for higher thresholds, the fractional distribution departs smoothly from $r^{-2}$. Bottom: Mollweide projection of multi-streams on the surface of the sphere.}
\label{fig:3}
\end{figure}

\begin{figure}
\begin{minipage}[t]{.99\linewidth}
  \centering\includegraphics[width=8.cm]{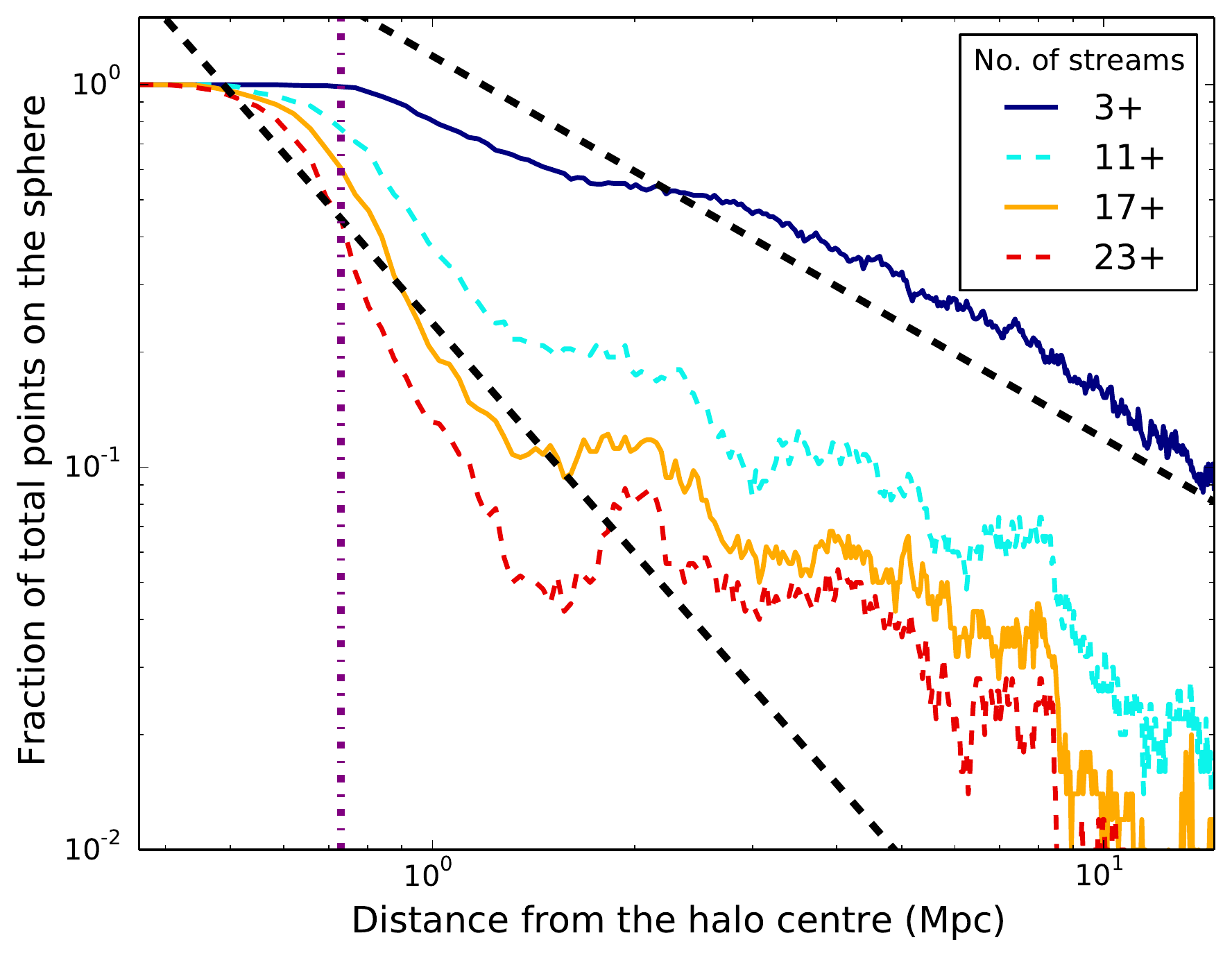}
\includegraphics[width=8.cm]{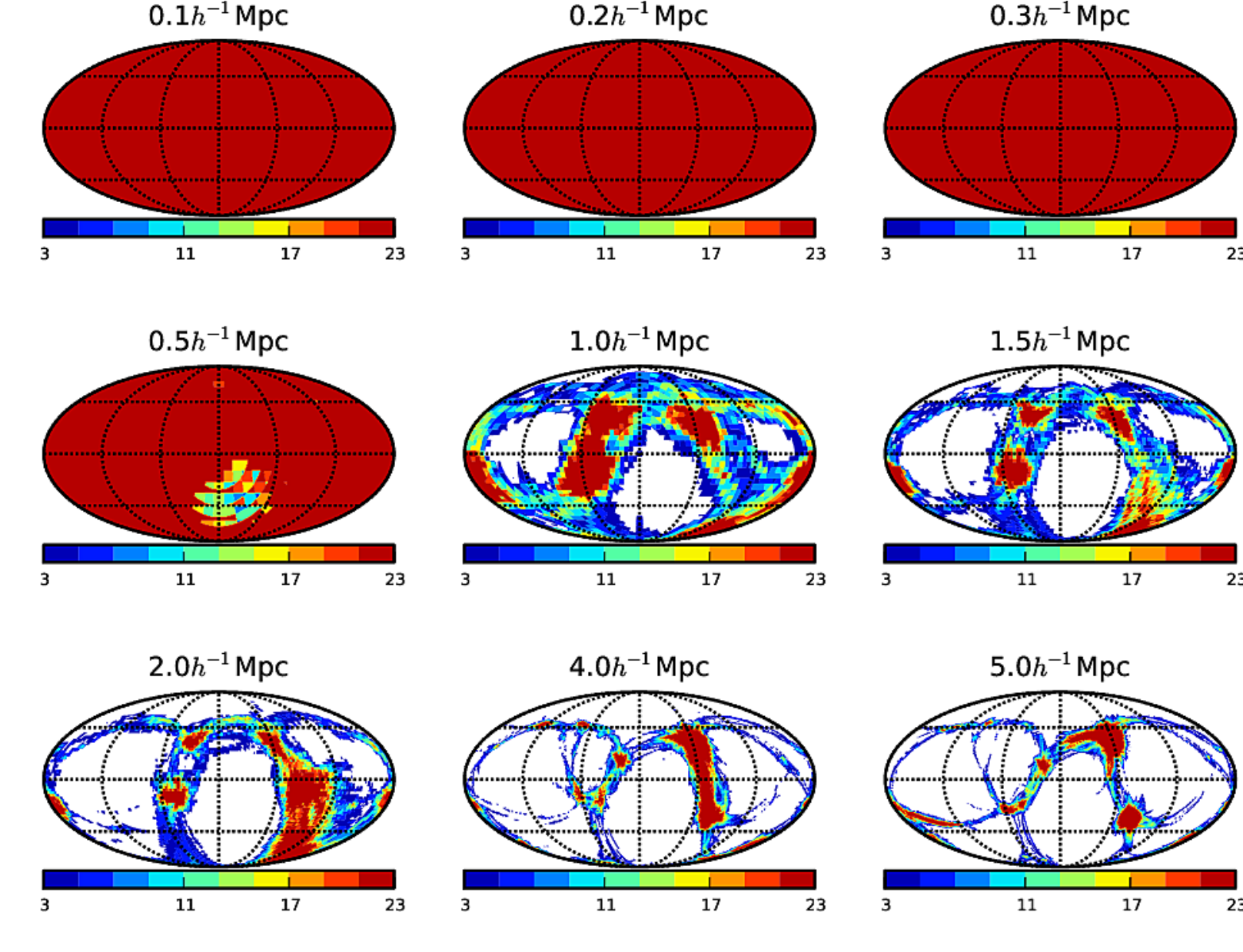} 
\end{minipage}\hfill
\caption{ Halo of mass $5.0 \times 10^{13} M_{\odot} $   and  FOF radius 0.7 $h^{-1}$ Mpc. Top: Fractional distribution of streams deviates from $r^{-1}$ and $r^{-2}$ scales
since the high stream filament passes along the surface of the sphere. Bottom: Filament passing through the surface is 
seen from 2 to 5 times the halo radius. }
\label{fig:79}
\end{figure}

\begin{figure}
\begin{minipage}[t]{.99\linewidth}
  \centering\includegraphics[width=8.cm]{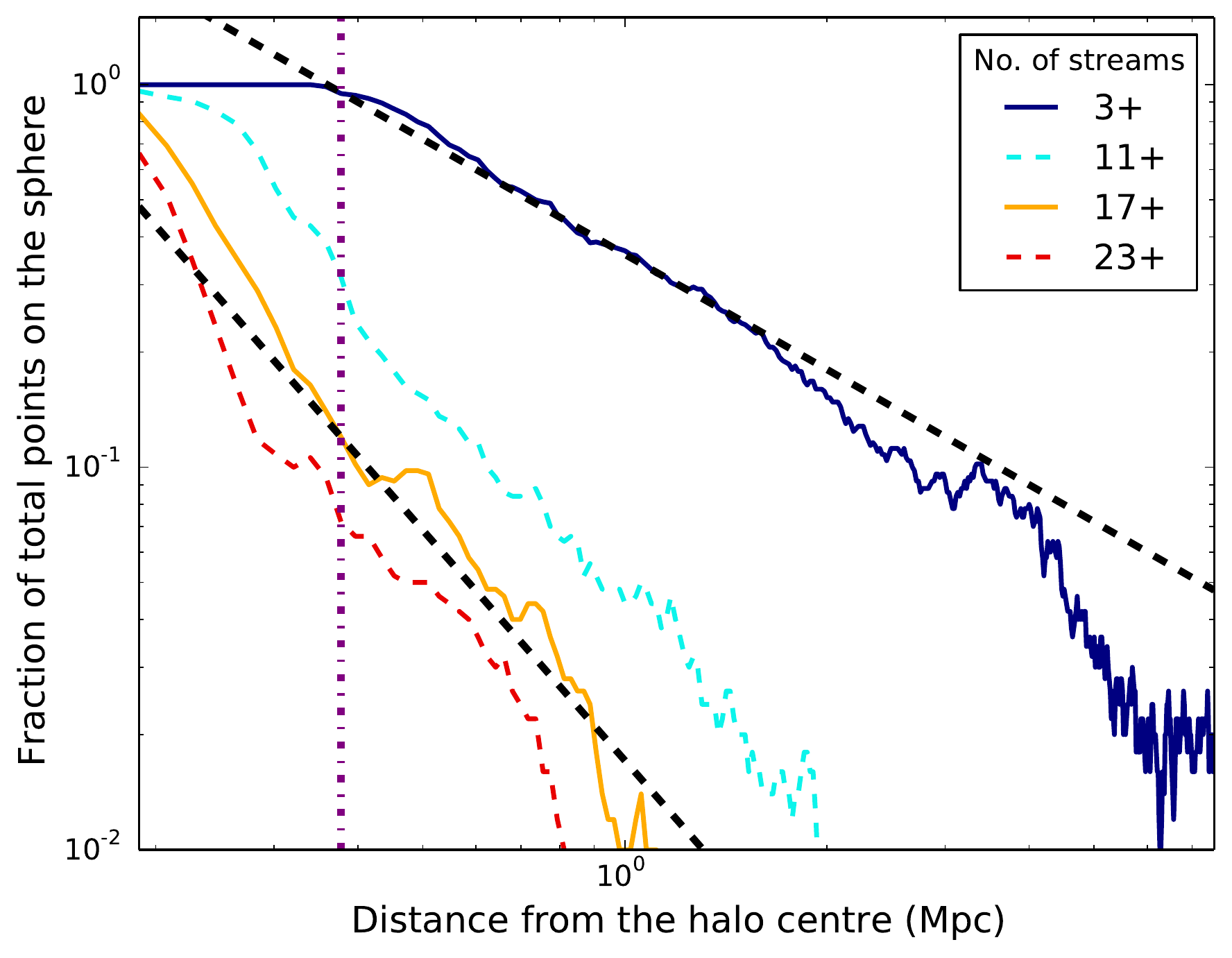}
\includegraphics[width=8.cm]{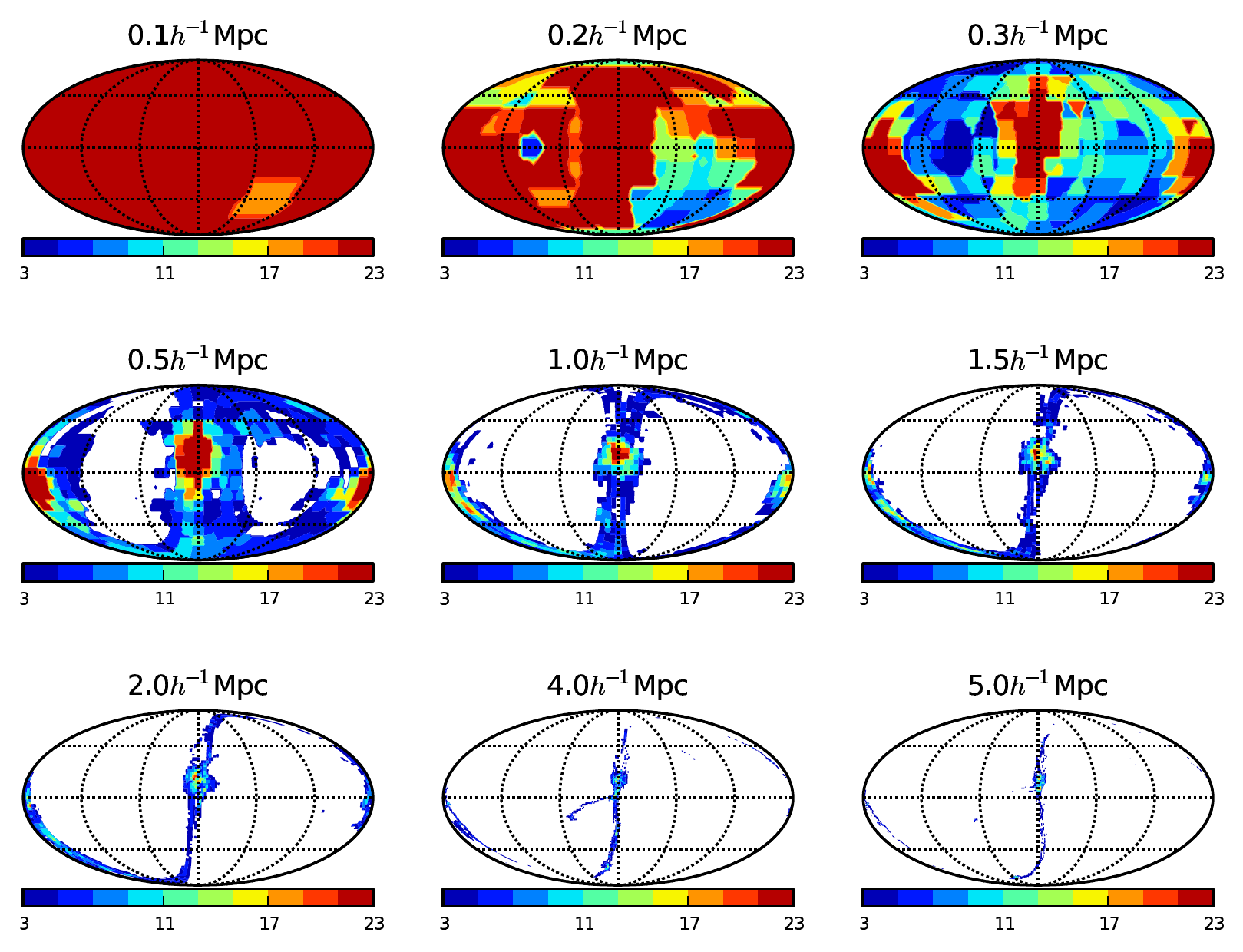} 
\end{minipage}\hfill
\caption{Halo of mass $7.0 \times 10^{12} M_{\odot} $  and  FOF radius 0.4 $h^{-1}$ Mpc. Top: All lines clearly scale between $r^{-1}$ and $r^{-2}$. Bottom: The filament is passing through the centre. It persists from radius of halo to 4 $h^{-1}$ Mpc. It is also surrounded by a single wall appearing as a line in the middle.}
\label{fig:706}
\end{figure}

\begin{figure}
\begin{minipage}[t]{.99\linewidth}
  \centering\includegraphics[width=8.cm]{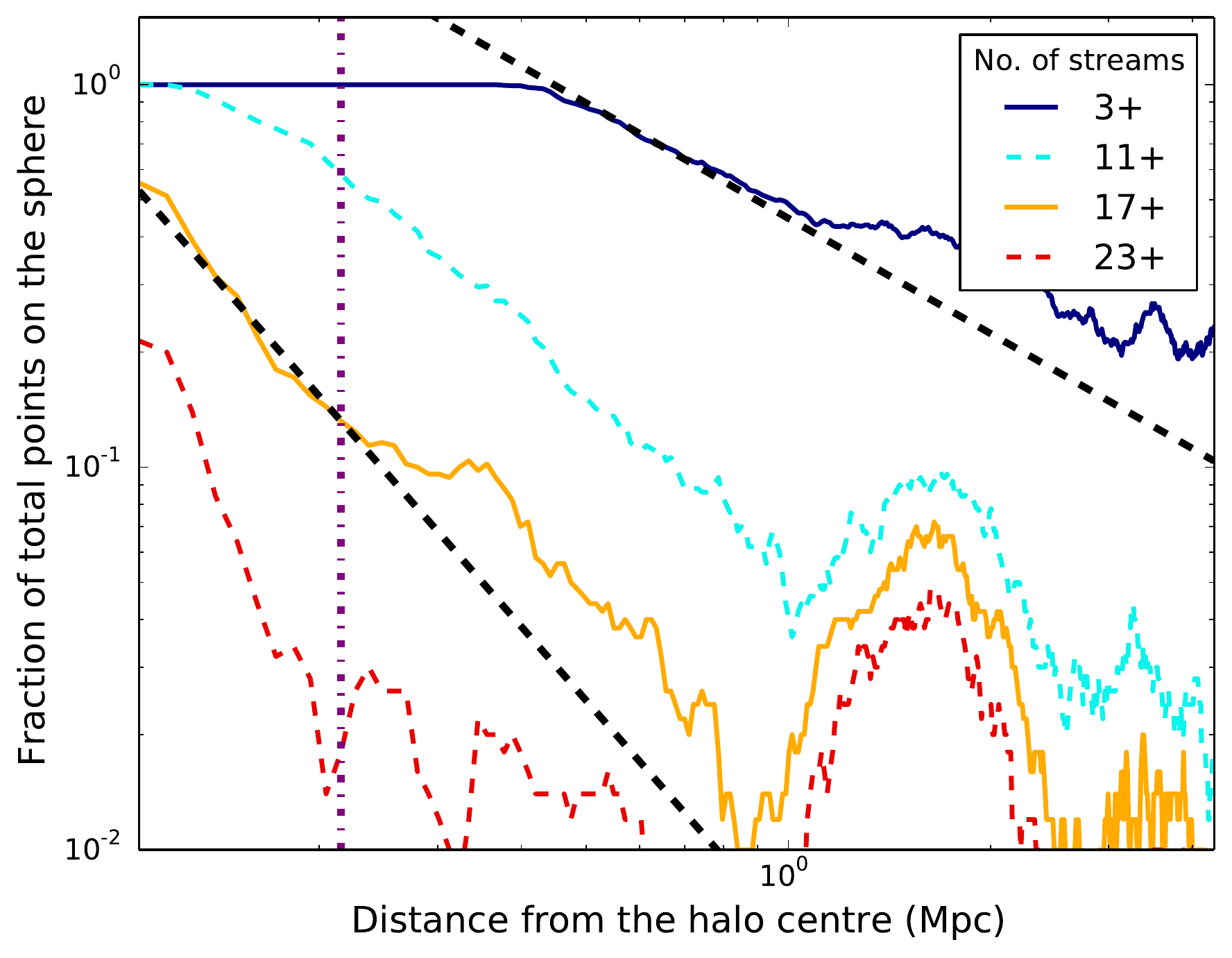}
\includegraphics[width=8.cm]{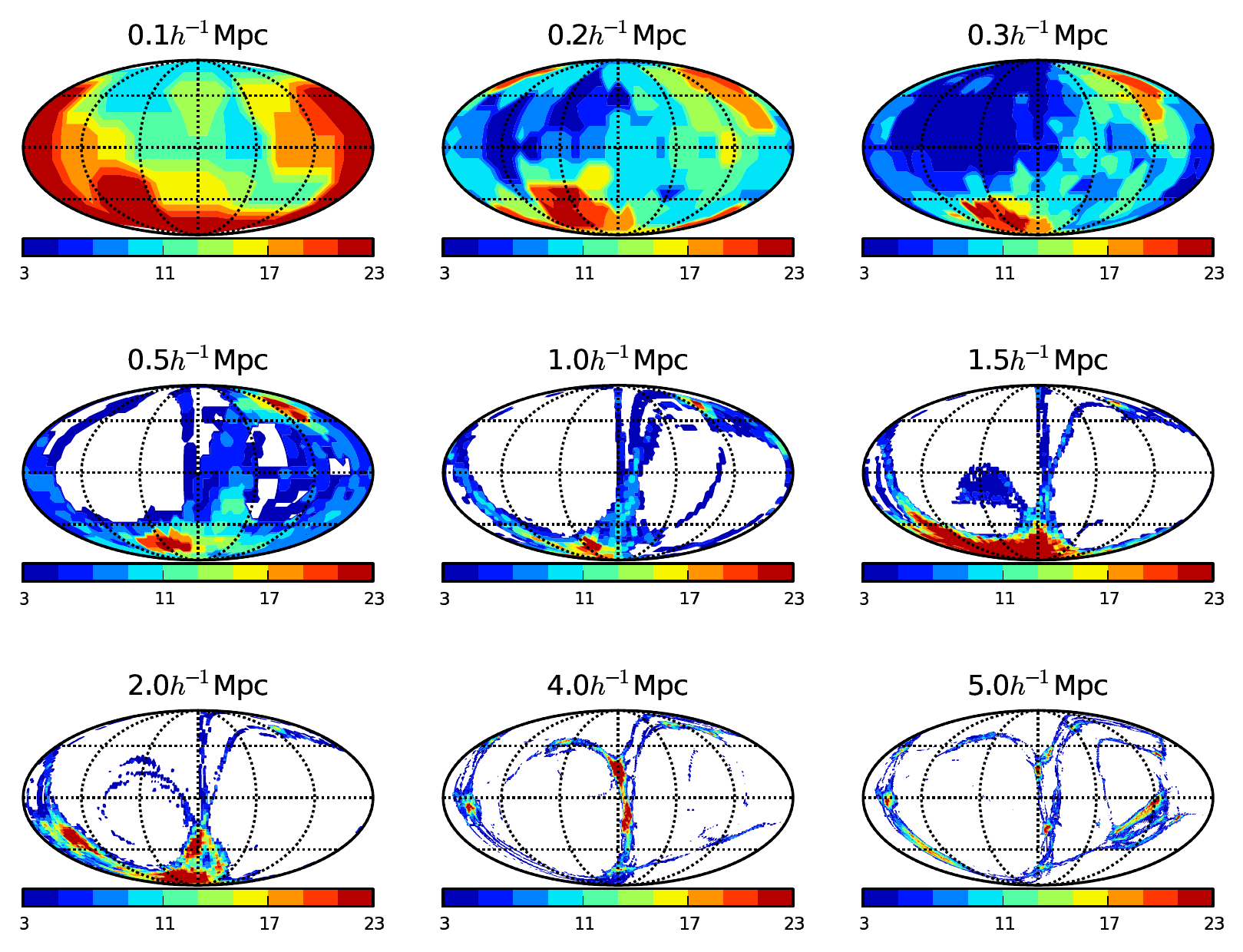}
\includegraphics[width=7.cm]{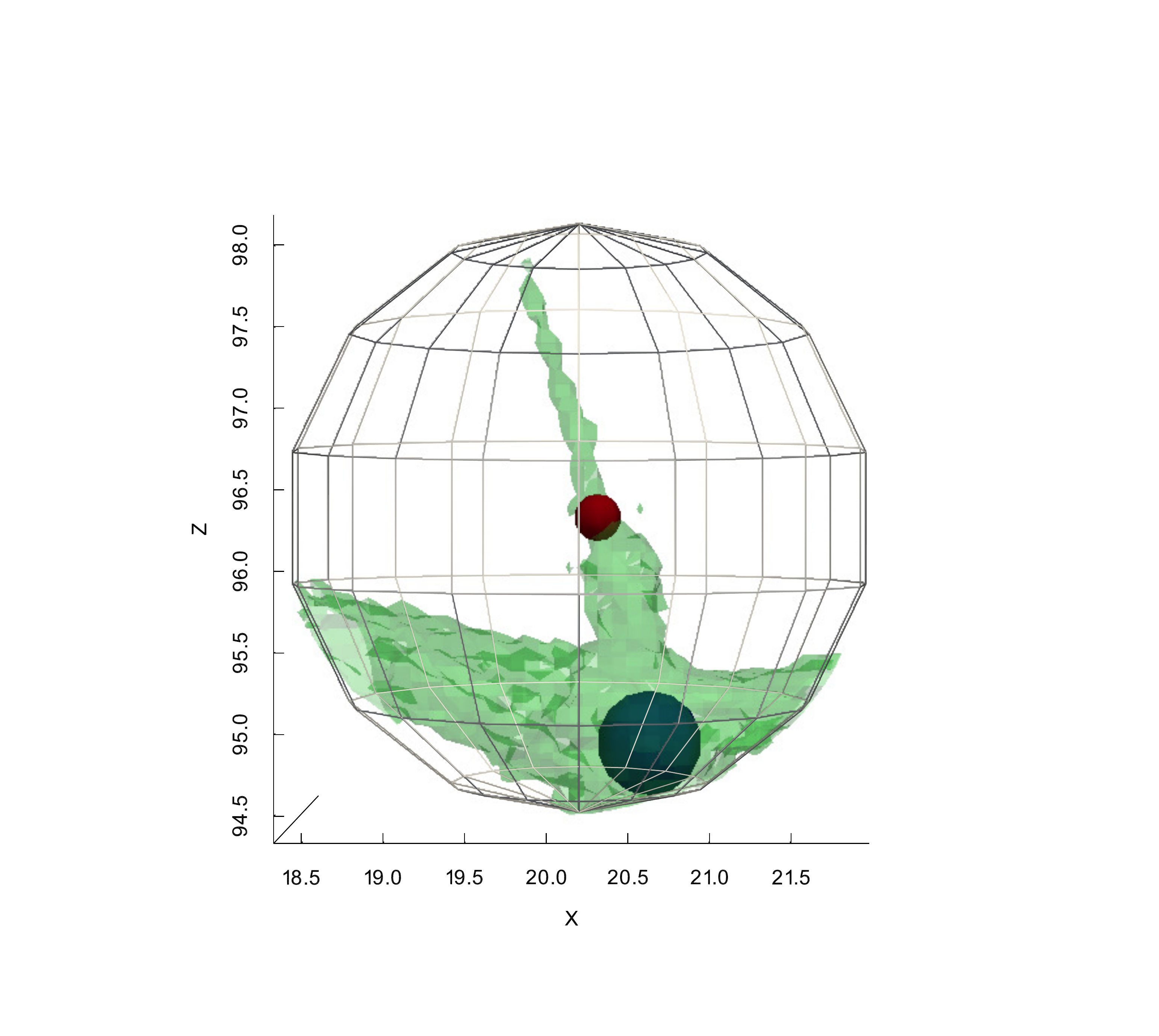}
\end{minipage}\hfill
\caption{Halo of mass $1.1 \times 10^{12} M_{\odot} $  and FOF radius 0.2 $h^{-1}$ Mpc (dotted-violet line). This halo just has 26 particles, hence the resolution is lesser than previous haloes for surfaces with low radii. Top: There is a bump in fraction of each of 3+ streams a little over 1 $h^{-1}$ Mpc. This is due to the presence of an additional halo nearby, as seen in the projections. The Mollweide projections from 0.4 $h^{-1}$ Mpc to 2 $h^{-1}$ Mpc have high stream flow regions near the lower surfaces of diagnostic spheres. Bottom: Corresponding FOF halo (red, at the centre) has a more massive neighbouring FOF halo (blue) within distance of 2 $h^{-1}$ Mpc. The 17+ stream regions (green) are increased around the neighbouring halo.} 
\label{fig:4322}
\end{figure}

The scaling of fraction of points with 3+ streams is closest to $r^{-1}$, where $r$ is the radius of the diagnostic sphere around the halo (Fig. ~\ref{fig:3} - ~\ref{fig:4322}; top figures). Since $r^{-1}$ variation is geometrically identical to that of a wall, it is identified as a flow region with 3+ stream flow. In this simulation the volume fraction of the web is dominated by 3-stream flows: $f_{vol}(3) \approx 0.04$ while
$\sum_5^{\infty} f_{vol}(n_{str}) \approx 0.02$.  

The deviation from the exact slope is expected, since assuming the filaments and walls as uniform cylinders 
and planes is rather crude. In the simulation, the filaments and walls have a far more complicated structure, where they branch out, and do not correspond to regular geometrical shapes. Detailed explanations for deviations are illustrated using Mollweide projections in the next section. 

Variation of multi-streams regions of 5+ to 17+ streams steadily changes from $r^{-1}$ to $r^{-2}$. This smooth transition implies that finding an exact cut-off of $n_{str}$ for  a filament is possible neither $n_{str}$ threshold nor by density. At 17+ stream regions scaling is closest to $r^{-2}$, the approximate filamentary geometry. In fact, $n_{str} = 19+$ regions also exhibit similar variation in spherical projections, but our choice of the threshold based on the lowest $n_{str}$ value that scales close to $r^{-2}$. Again, unlike the threshold for voids and walls, the threshold for filaments and haloes are not unambiguous. Our seemingly arbitrary choice of definition of filaments as regions with 17+ streams (in Section \ref{sec:global}, Fig. \ref{fig:full} and Fig. \ref{fig:4view}) was motivated by this observation. 
Thus projections on a diagnostic sphere is a convenient tool for classifying regions in the simulation as belonging to void, wall, filament or a halo. 

For illustrations, we have picked 4 haloes from different mass ranges: $3.7 \times 10^{14}M_{\odot} $,  $5.0 \times 10^{13} M_{\odot} $ , $7.0 \times 10^{12} M_{\odot} $ and $1.1 \times 10^{12} M_{\odot} $ from the simulation box of 100 $h^{-1}$ Mpc length and $128^3$ particles. Multi-stream field with a high refinement factor of 8 is calculated for a greater resolution on scales of the halo volume. Diagnostic spheres of radii 0.1 $h^{-1}$ Mpc to 5 $h^{-1}$ Mpc are drawn for each of these haloes (Fig. ~\ref{fig:3} - ~\ref{fig:4322}; bottom figures), with the multi-stream field projected onto the surface. In the Mollweide projections of these spheres, the white space refers to single-stream voids. For the largest halo (Fig. ~\ref{fig:3}) with FOF radii 1.2 $h^{-1}$ Mpc, the voids already appear in sphere of radius 1.5 $h^{-1}$ Mpc and in the smaller haloes (Fig. \ref{fig:4322}) it appears as early as 0.5 $h^{-1}$ Mpc.  

Up to 1 $h^{-1}$ Mpc from halo center of the largest halo, the surfaces are uniformly covered with high number of streams (red, 17+). This shows that the most non-linear regions are close to centres of haloes. A similar trend is seen for the halo of radius 0.7 $h^{-1}$ Mpc (Fig. ~\ref{fig:79}). However, for smaller haloes (Fig. \ref{fig:706} and \ref{fig:4322}) lower number of streams (even the wall forming 3+ streams; blue) start occupying the spherical surface at radii lesser than FOF-radius. In the case of the smallest halo of $1.1 \times 10^{12} M_{\odot} $ mass, the 17+ streams are seen at scales as low as 0.1 $h^{-1}$ Mpc. The distribution of multi-streams on the surface seems do not have a symmetry of any kind , signifying a complex morphology of the web in the vicinity of the haloes. Regions with 5+ to 15+ streams form structures intermediary to filament-like and wall-like behaviour, as seen by scaling of fraction of total points on the space with distance from halo center.

Halo environment at distance over twice the FOF radius reveals interesting morphological features. The walls intersect the sphere, and in the 
Mollweide projections, appear like a thin strip. 
We also note that a filament oriented tangentially to the diagnostic surface may occasionally appear as a strip too (like in Fig. ~\ref{fig:79}, 
see the corresponding discrepancy in fraction of streams), but upon inspecting the spheres at various radii, we can clearly identify the persisting line-like structures, and they correspond to the walls. Similarly, a filament is projected as a compact patch structure, which occurs due to an intersection of a cylinder-like geometry with the spherical surfaces. It is clearly observed at the distance of 4 - 5 $h^{-1}$ Mpc in Fig. \ref{fig:3} and in between 0.5 - 5.0 $h^{-1}$ Mpc in Fig. \ref{fig:706}.
	
Hence we conclude that the 3+ stream regions constitute predominantly walls and the regions with 17+ streams correspond mostly to filaments. The higher $n_{str}$ shells must be surrounded by the layers with lower $n_{str}$. Thus, the filaments are within the walls, and do not exist independently.  
We remind that the radius of diagnostic sphere varies from 0.1 - 5 $h^{-1}$ Mpc, whereas the Mollweide projections shown here are of the same size. Hence the walls and filaments appear more narrow and smaller in larger spheres due to zooming-out effects. 
 In some cases (Fig. \ref{fig:79}, ~\ref{fig:4322}), the Mollweide projections display the walls  and filaments as a complicated network 
 with patches of high number of streams. 

The high peak shown by the curves corresponding to the numbers of streams from 11+ to 17+  in the top panel of Fig. ~\ref{fig:4322}
is mostly due to the presence of another halo nearby ( seen at bottom of Mollweide projection of 1 - 2 $h^{-1}$ Mpc). Fig. \ref{fig:79} also has a deviation from usual scaling, and this due to the intricate shape of 17+ stream filament, which appears to be branching out after 1 $h^{-1}$ Mpc. 

Generally the transitions from haloes to filaments then to walls and finally to voids appear to be rather smooth.
However occasionally sharp features as the one seen in Fig. ~\ref{fig:4322} may emerge when the diagnostic sphere hits a neighbouring halo.


 \begin{figure}
\begin{minipage}[t]{.99\linewidth}
  \centering\includegraphics[width=8.cm]{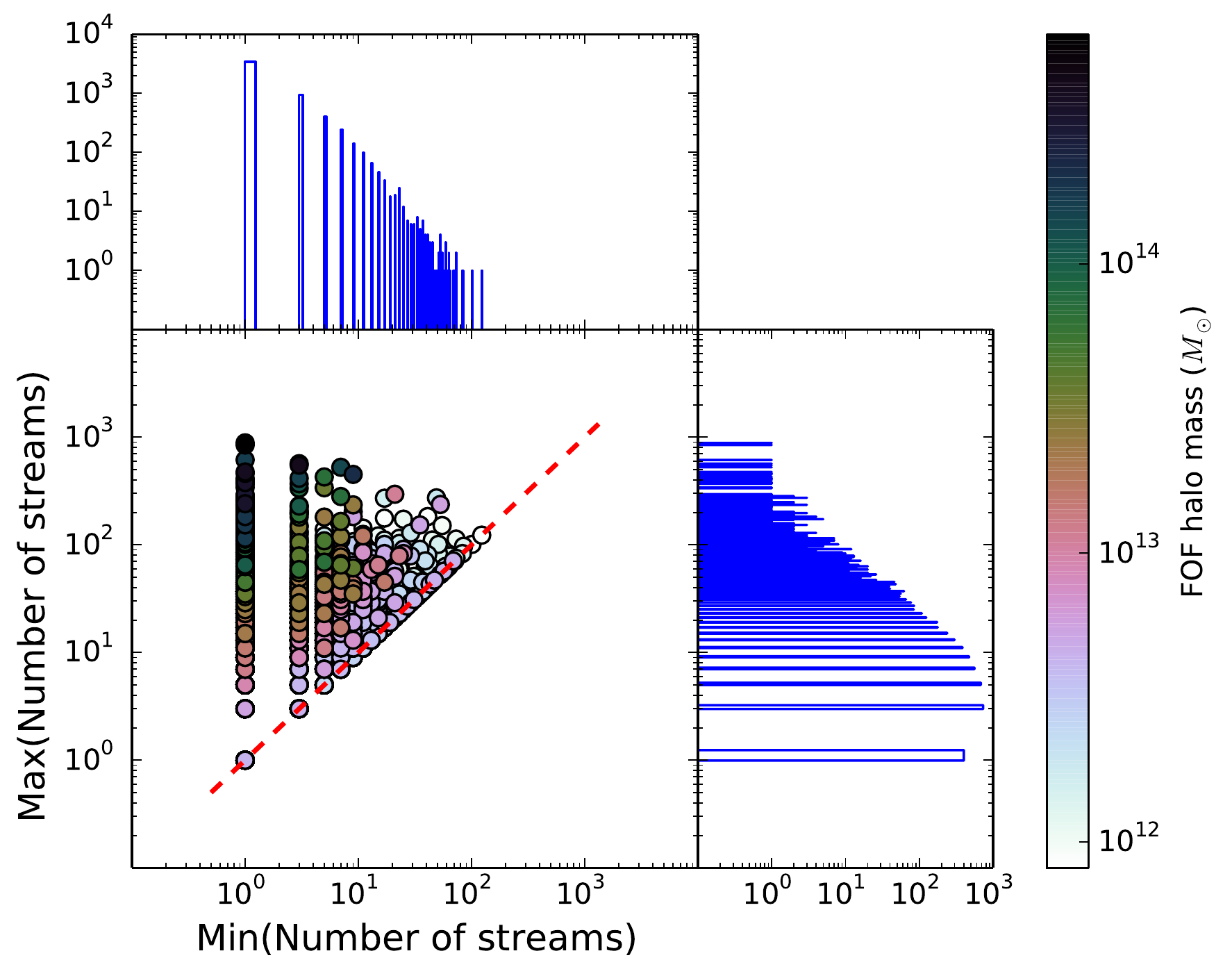}
\includegraphics[width=8.cm]{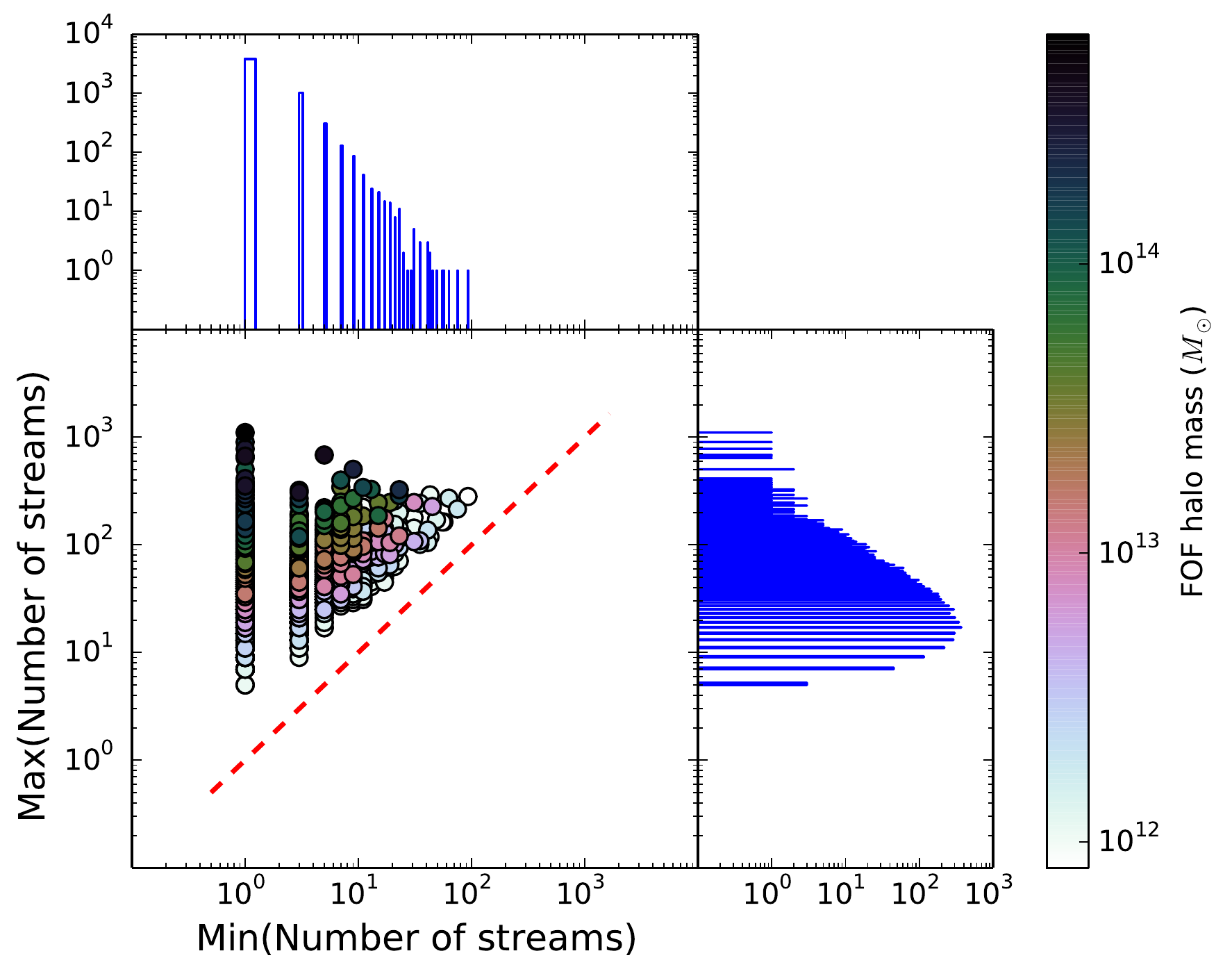} 

\end{minipage}\hfill
\caption{Scatter plot of minimum and maximum number of streams on the surface of FOF radius around all haloes in FOF catalogue. For the analysis, we use a total of 5521 haloes that are identified using the FOF technique with linking length, $b=0.2$. Since several of the haloes coincide, distributions of number of haloes corresponding to minimum and maximum $n_{str}$ on their FOF radii are shown above and beside the scatter plot respectively. Top: Full box of $L =100 ~h^{-1}$ Mpc and $N = 128^3$ (i.e. $L/N = 0.78 \,h^{-1}$), with a low refinement factor of 1 is utilized for multi-stream field calculation. Bottom: Same simulation box; multi-stream field calculated with higher refinement factor of 8.}
\label{fig:minmax128}
\end{figure}

 \begin{figure}
\begin{minipage}[t]{.99\linewidth}
  \centering\includegraphics[width=8.cm]{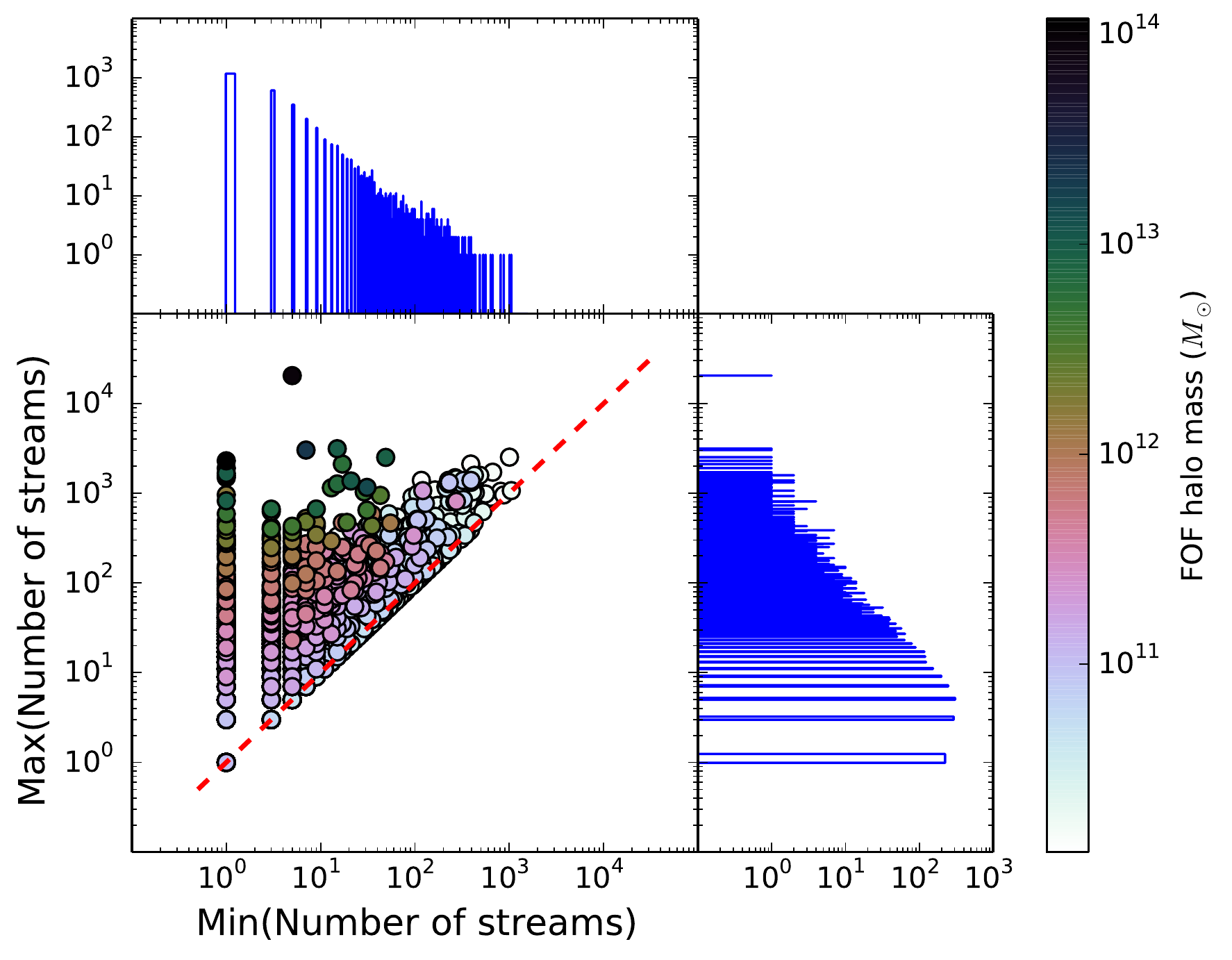}
\includegraphics[width=8.cm]{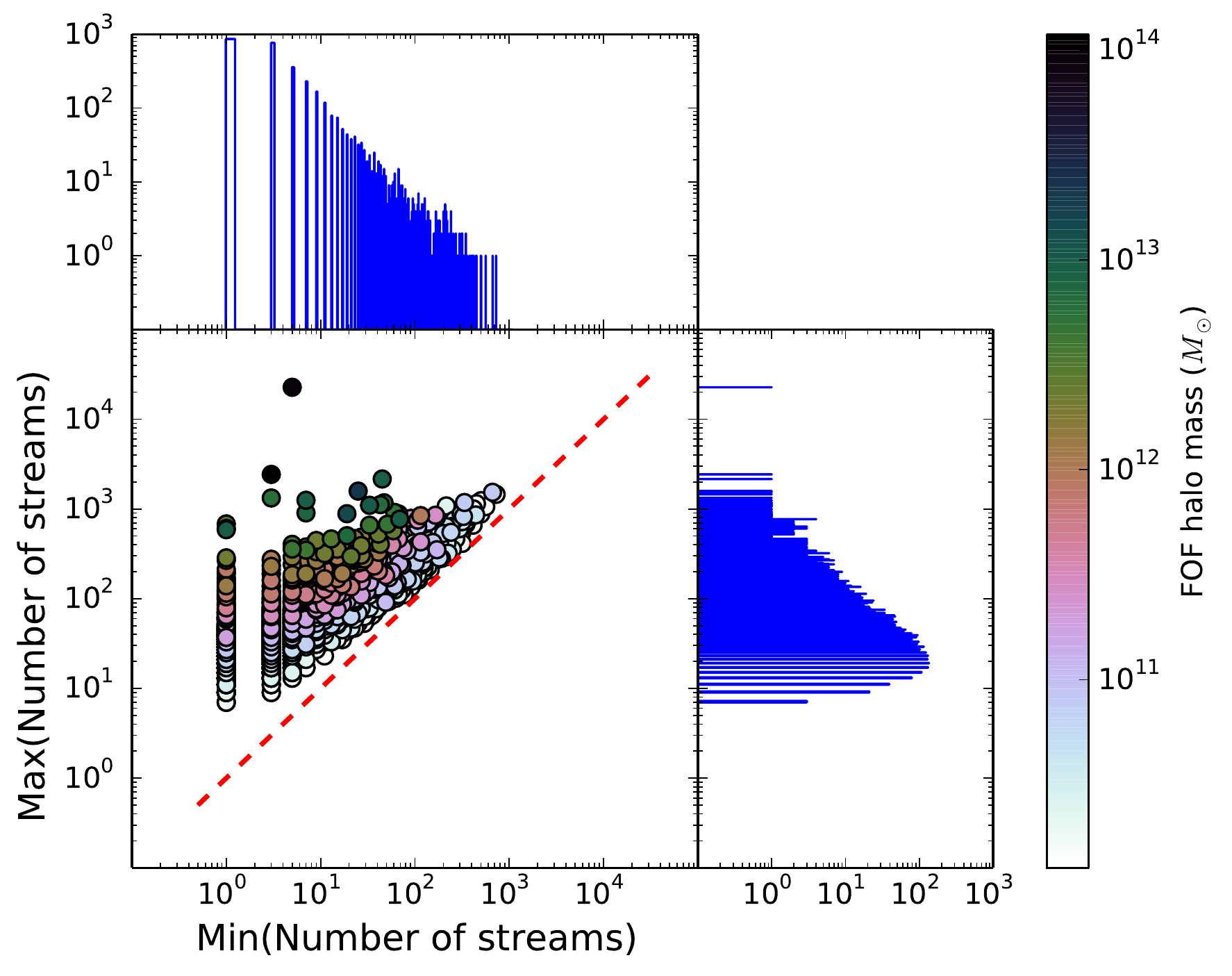} 
\end{minipage}\hfill
\caption{Scatter plot of minimum and maximum number of streams on the surface of FOF radius around all haloes in FOF catalogue. Distributions of number of haloes corresponding to minimum and maximum $n_{str}$ on their FOF radii are shown above and beside the scatter plot respectively. Top: In the simulation box of $L =100 ~h^{-1}$ Mpc and $N = 512^3$ (i.e. $L/N = 0.19 \,h^{-1}$), multi-stream field is calculated on a smaller slice of $ 25 ~h^{-1}$ Mpc is  with a refinement factor of 1. Multi-stream field is projected onto surfaces of 3448 FOF haloes within this small box. Bottom: Same simulation box; the multi-stream field calculated on the the same small box, but with a refinement factor of 8.}
\label{fig:minmax512}
\end{figure}

The friends-of-friends analysis identifies haloes as spherical structures. The distribution of multi-streams projected onto these surfaces of the FOF haloes can be utilized for a statistical analysis of the haloes (Fig. \ref{fig:minmax128} and \ref{fig:minmax512}). We have utilised FOF catalogues of haloes more than 20 particles found at linking length $ b= 0.2$.  The $n_{str}$ ranging from as low as 1 to higher than 10$^3$ are seen on the halo surfaces. Haloes which have the minimum $n_{str} = 1$ are in contact with the void. However, if the maximum $n_{str}$ is also 1, then the halo is completely within the void. In calculations with low refinement factor of 1, only 7.3$\%$ (for  $L/N = 0.78 \,h^{-1}$) and 6.5$\%$ (for  $L/N = 0.19 \,h^{-1}$) of the haloes are completely within single streaming voids. This is solely due to low resolution of multi-stream field, since none of the FOF haloes are found in high resolution multi-stream calculations on both $L/N = 0.78 \,h^{-1}$ and $L/N = 0.19 \,h^{-1}$ boxes. At high refinement factors, none of the haloes are entirely embedded in a region with just one multi-stream value (i.e., max($n_{str}$) = min($n_{str}$), along the dotted-red lines in Fig. \ref{fig:minmax128} and \ref{fig:minmax512}). However, there are significant number of haloes whose FOF surfaces are in contact with the void region: in calculations with refinement factor of 8, 62$\%$ of the haloes in $L/N = 0.78 \,h^{-1}$ and 34$\%$ of the haloes in $L/N = 0.19 \,h^{-1}$ are in contact with void on their FOF radii. Rest of the haloes are completely within non-void regions.

Statistical analysis of FOF haloes in Fig. \ref{fig:minmax128} and \ref{fig:minmax512} show that massive haloes tend to have low min($n_{str}$) and high max($n_{str}$), hence a very diverse multi-stream environment on their spherical surface. The heuristic  multi-stream threshold for haloes mentioned in Section \ref{sec:global} results in virialized haloes with a constant $n_{str}$ value. These halo surfaces far from sphere (see Fig. \ref{fig:4view}), whereas, the FOF surfaces are spherical and have a large range of number of streams on their surfaces. The probability distribution function of number of FOF haloes has an approximately exponential tail monotonically decreasing with with min($n_{str}$).


\section{Summary}
In this paper, we explore for the first time, the multi-stream environment of dark matter haloes in cosmological N-body simulations.
The visualization and quantitative characterization of generic non-linear fields in three-dimensional space represent a serious
challenge from both conceptual and computational points of view. The complexity of the problem requires diverse tools
for analysing the results of cosmological simulations as well as galaxy catalogues.

This study is different from the most previous works in a few aspects.
Firstly, we consider the representation of the cosmic web in the form of  a multi-stream field rather  than a density field.
The multi-stream field contains a different information about the web than the density and velocity fields and thus represents 
a complimentary characterization of the web revealing new dynamical features of the web 
(\citealt{Shandarin:11}, ~\citealt{Shandarin_etal:12},~\citealt{Abel_etal:12}).  Secondly, for computing the multi-stream field we use 
the tessellation of three-dimensional Lagrangian sub-manifold ${\bf x} = {\bf x}({\bf q}, z=0)$ in six-dimensional $({\bf x},{\bf q})$ space
which allows to significantly increase the spatial resolution (\citealt{Shandarin_etal:12},~\citealt{Abel_etal:12}). The Lagrangian sub-manifold is more convenient since
{\bf x}  is a single-valued function of {\bf q} at any stage including a highly non-linear regime while 
the phase space sheet projected on  {\bf x}- or {\bf v}-space is not. If the initial state of the simulation is represented by a uniform three-dimensional mesh, then storing the Lagrangian sub-manifold does not require additional space for Lagrangian coordinates. And thirdly, in the study of the multi-stream environment of dark matter haloes we use the Mollweide projection of the multi-stream field 
computed on a set of diagnostic spherical surfaces centred at the FOF haloes and  having radii from 0.1 $h^{-1}$Mpc to 5 $h^{-1}$Mpc.

Most of the results are obtained for a simulation in $L=100  h^{-1}$Mpc box with $N = 128$  particles along each axes although we report
some of the results for the simulations in 100  $h^{-1}$Mpc box with 256$^3$ and 512$^3$ particles as well as  in 200  $h^{-1}$Mpc
box with 128$^3$, 256$^3$  and 512$^3$ particles.

Using the tessellation of the three-dimensional Lagrangian sub-manifold 
${\bf x} = {\bf x} ({\bf q},t)$  \citep{Shandarin_etal:12}, we compute the multi-stream field i.e. 
the number of streams  on a regular grid in the configuration ${\bf x}$-space, $n_{str}({\bf x})$ for
estimating global parameters or on selected set of points in the study of the haloes environments.

The multi-stream field takes odd whole numbers everywhere except at a set of points of measure zero where it takes positive even whole numbers.
This property is very useful for debugging the numerical code.

 The multi-stream field allows one to define physical voids as the regions with $n_{str} =1$. 
 The rest of space with $n_{str} \ge 3$ can be called the non-void or web.
This division of the space into two parts is unique and physically motivated: no object can form before shell crossing happens. 
It is worth emphasizing that the division of space into voids and web is based on the local parameter, the number of streams at a single point (\citealt{Shandarin_etal:12}, \citealt{Abel_etal:12}). \citet{Flack_etal:12} and \citet{Flack_Neyrinck:14} defined the web as a set of particles that experienced the 'flip-flop' at least once along any axis. We discussed potential problems with this definition in the beginning of Section \ref{sec:global}.

   
 The further division of the web into walls, filaments and haloes is not straightforward although haloes can be defined using dynamical parameters
 related to the requirement of virialization of haloes. One of the simplest is the famous density threshold $\rho/\bar{\rho} \approx 200$. 
 Identifying filaments and walls is significantly more tricky (see e.g. \citealt{Hahn_etal:07}, \citealt{Forero_etal:09}, 
  \citealt{Aragon_etal:10}, \citealt{Cautun_etal:14}, \citealt{Flack_Neyrinck:14}) and require non-local parameters.
  
The large part of walls can also be identified locally since the regions where $n_{str} = 3$ can be neither filaments nor haloes. 
For instance, in the simulation  100  $h^{-1}$Mpc box with 128$^3$  particles the web occupies about 6\% of the volume, the three-stream flow
regions occupy about 4\% and the rest of the web remaining 2\% of the total volume.

In this study we introduced an empirical statistical criteria which very crudely distinguish wall, filament and haloes. 
We have found empirically that  in  the studied simulation the transition from wall points to filament points takes place approximately at $ 5 \le n_{str} \lesssim 15$. Using the virial over-density threshold of 200 in Eq.~\ref{eq:den}, we have also estimated that the haloes correspond to the regions with $n_{str} \gtrsim 90$.
Thus, the transition from filament to haloes takes place in the range $ 17 \le n_{str} \lesssim 90$.
The above critical values for transition from walls to filaments and from filaments to haloes were shown to be approximately correct
for  the simulation with $L/N = 0.78 h^{-1}$ Mpc. This technique can be also applied to simulations with different $L/N$ ratio and multi-stream grid of different refinement factors but the classification based on the threshold applied locally will remain only a very crude estimator. A more sophisticated morphological analysis will require non-local geometric and topological methods, which is beyond the scope of this paper.
 
 \begin{table*}
  \caption{Comparison of the volume and mass fraction of the elements of the cosmic web between our analysis and \citet{Flack_Neyrinck:14}. $ L/N = 0.78 h^{-1}$ Mpc for simulations used in both techniques. We use a refinement factor of 8 for the multi-stream grid. The mean density is given in units of the mean density of the universe. }
\begin{tabular}{|l|r|r|r|r|r|r|r|r|}
\hline
\multicolumn{1}{|c|}{} & \multicolumn{4}{l|}{Multi-stream analysis (This work)} & \multicolumn{4}{c|}{ORIGAMI \citep{Flack_Neyrinck:14}}    \\ \hline
                       & Voids & Walls & Filaments & Haloes  & Voids & Walls & Filaments & Haloes \\ \hline
Volume Fraction (\%)                 & $93$   & $7$    & $< 1$      & $< 0.1$ & $84$   & $12$   & $3$        & $< 1$  \\ \hline
Mass Fraction   (\%)               & $32$   & $35$   & $17$       & $14$    & $26$   & $19$   & $19$       & $35$   \\ \hline
Mean density                    & $0.34$ & $5$    & $>17$      & $>140$  & $0.31$ & $2.2$  & $6$        & $>35$  \\ \hline
\end{tabular}
 \label{tab:Compare}
\end{table*}

We have found that the volume and mass fractions in the voids  are approximately V.F./M.F. = 96/76, 93/32  and 88/24, where each number
is the percentage, for the simulations with  $L/N = 1.56 \,h^{-1}$, $0.78 \,h^{-1}$  and $0.19 \,h^{-1}$ Mpc respectively. As the ratio $L/N$ 
gets smaller both volume and mass fractions in  voids monotonically decrease.  
This is fairly consistent with the results of \citet{Flack_Neyrinck:14}, considering the differences between our numerical methods. We compare the fractions of the volumes and masses
in other components of the web in Table \ref{tab:Compare}, and the results are in a good qualitative agreement.
Our estimates systematically higher for both volume and mass fractions for voids and thus systematically lower for the web. 
One general conclusion seems to be obvious: the web defined by the multi-stream flows is about  (100\% - 84\%)/(100\%-93\%)=2.3  times  thinner than that defined by the ORIGAMI  method.


In conclusion we would like  to outline the major aspects of the web revealed by the study of the multi-stream field.
The multi-stream field is a fundamental attribute of the structures formed in cold collision-less dark matter. Its properties
are of great importance for the detecting dark matter directly in a laboratory setting or indirectly via astronomical observations.
The dark matter web described by a multi-stream field represents a nested structure, consisting of layers with increasing number of streams.
The number of streams are odd integers almost everywhere except on caustics where they are even integers.
The caustics  occupy infinitesimal volume. The most of the volume is occupied by one stream flow regions which
are dark matter voids. The regions with three streams are the regions occupying the second largest volume.
They form very thin membrane type structures (often referred to as walls or pancakes) most of which are connected 
in one huge connected formation. The three-stream regions form the external shell of the web. All other structures
filaments and haloes are within the three-stream shell.
 The membranes are attached to each others by the filaments which locally consist of regions with higher number of streams  than
the neighbouring membranes. The filaments form the framework or a skeleton of the dark matter web.  Similar to a real skeleton, the filamental structure has joints where most of the the dark matter haloes are located. The haloes are the local peaks of the multi-stream field.

\section*{Acknowledgements}
The authors are grateful to Gustavo Yepes and Noam Libeskind for providing the simulations used in this study and the Sir John Templeton Foundation for the support.The authors would also like to thank the anonymous referee for careful review and useful suggestions.

\bibliographystyle{mn2e}
\bibliography{bibliography}

\end{document}